\documentclass{emulateapj}
\submitted{Accepted for publication in \apj}
\usepackage{apjfonts}
\usepackage{amsmath}

\newcommand\lsim{\mathrel{\rlap{\lower4pt\hbox{\hskip1pt$\sim$}}
        \raise1pt\hbox{$<$}}}
\newcommand\gsim{\mathrel{\rlap{\lower4pt\hbox{\hskip1pt$\sim$}}
        \raise1pt\hbox{$>$}}}
\newcommand{\D}{\mathrm{d}}
\newcommand{\km}{\,\mathrm{km}}
\newcommand{\yr}{\,\mathrm{yr}}
\newcommand{\days}{\,\mathrm{day}}
\newcommand{\hr}{\,\mathrm{hr}}

\newcommand{\Hz}{\,\mathrm{Hz}}
\newcommand{\mHz}{\,\mathrm{mHz}}
\newcommand{\Gpc}{\,\mathrm{Gpc}}
\newcommand{\Mpc}{\,\mathrm{Mpc}}

\newcommand{\pc}{\,\mathrm{pc}}
\newcommand{\gc}{\mathrm{gc}}
\newcommand{\gal}{\mathrm{gal}}
\newcommand{\hn}{S_\mathrm{n}}
\newcommand{\CO}{\mathrm{CO}}
\newcommand{\NS}{\mathrm{NS}}
\newcommand{\BH}{\mathrm{BH}}
\newcommand{\vir}{\mathrm{vir}}
\newcommand{\rel}{\mathrm{rel}}
\newcommand{\Mchirp}{\mathcal{M}}
\newcommand{\Msun}{\mathrm{M}_{\odot}}
\newcommand{\G}{\mathrm{G}}
\newcommand{\cl}{\mathrm{c}}

\newcommand{\dL}{d_\mathrm{L}}
\newcommand{\AU}{\,\mathrm{AU}}
\newcommand{\I}{\mathrm{I}}
\newcommand{\II}{\mathrm{II}}

\shorttitle{Detection rate estimates of parabolic encounters of
stellar black holes} \shortauthors{Kocsis, G\'asp\'ar, \& M\'arka}

\begin{document}

\title{Detection rate estimates of gravity-waves emitted
during parabolic encounters of stellar black holes in globular
clusters}

\author{Bence Kocsis}
\affil{Institute of Physics, E\"otv\"os University, P\'azm\'any P.
s. 1/A, 1117 Budapest, Hungary; bkocsis@complex.elte.hu}
\author{Merse El\H{o}d G\'asp\'ar}
\affil{Institute of Physics, E\"otv\"os University, P\'azm\'any P.
s. 1/A, 1117 Budapest, Hungary; merse@complex.elte.hu}
\author{Szabolcs M\'arka}
\affil{Department of Physics, Columbia University, 550 West 120th
Street, New York, NY 10027; smarka@phys.columbia.edu}

\begin{abstract}
The rapid advance of gravitational-wave (GW) detector facilities
makes it very important to estimate the event rates of possible
detection candidates. We consider an additional possibility of GW
bursts produced during unbound orbits of stellar mass compact
objects. We estimate the rate of successful detections for specific
detectors: the initial Laser Interferometric Gravitational-Wave
Observatory (InLIGO), the French-Italian gravitational-wave antenna
VIRGO, the near-future Advanced-LIGO (AdLIGO), the space-based {\it
Laser Interferometric Space Antenna} ({\it LISA}), and the {\it Next
Generation LISA} ({\it NGLISA}). The dominant contribution among
unbound orbits that have GW frequencies in the sensitive band of the
detectors correspond to near-parabolic encounters (PEs) within
globular clusters (GCs). Simple GC models are constructed to account
for the compact object mass function, mass segregation, number
density distribution, and velocity distribution. We calculate
encounters both classically and account for general relativistic
corrections by extrapolating the results for infinite mass ratios.
We also include the cosmological redshift of waveforms and event
rates.
We find that typical PEs with masses $m_1=m_2=40\Msun$ are
detectable with matched filtering over a signal to noise ratio
$S/N=5$ within a distance $\dL\sim 200\Mpc$ for InLIGO and VIRGO,
$z=1$ for AdLIGO, $0.4\Mpc$ for {\it LISA}, and $1\Gpc$ for {\it
NGLISA}.
We estimate single datastream detection rates of $5.5\times
10^{-5}\yr^{-1}$ for InLIGO, $7.2\times 10^{-5}\yr^{-1}$ for VIRGO,
$0.063\yr^{-1}$ for AdLIGO, $2.9\times10^{-6}\yr^{-1}$ for {\it
LISA}, and $1.0\yr^{-1}$ for {\it NGLISA}, for reasonably
conservative assumptions.
These estimates are subject to uncertainties in the GC parameters,
most importantly the total number and mass-distribution of black
holes (BHs) in the cluster core. In reasonably optimistic cases, we
get $\gsim 1$ detections for AdLIGO per year. We can expect that a
coincident analysis using multiple detectors and accounting for GW
recoil capture significantly increases the detection rates.
We give ready-to-use formulas to recalculate the estimates when
these input parameters become better-determined. In addition, we
provide the partial detection rates for various masses.
The regular detection of GWs during PEs would provide a unique
observational probe for constraining the stellar BH mass function of
dense clusters.
\end{abstract}

\keywords{gravitational waves -- black holes -- globular clusters:
general}

\section{Introduction}\label{sec:intro}

Interferometric gravitational-wave (GW) detectors LIGO, GEO, TAMA,
and VIRGO are searching for GW signals with unprecedented
sensitivity \citep{hmbh01,LIGO05a,a05a,a05b,geo05,tama05,virgo05}.
For LIGO, the noise levels are already reaching the goal level
necessary for the detection of the strongest signals. It is very
important to analyze the detection capabilities of these detectors
and to estimate the rates of potentially detectable GW signals.
There is already a considerable list of possible detection
candidates \citep[for a review see][]{ct02}: the inspiral of neutron
star (NS) or black hole (BH) binaries, the tidal disruption of NS by
BH in NS--BH binaries, BH--BH merger and ringdown, low-mass X-ray
binaries, pulsars, centrifugally hung-up proto neutron stars in
white dwarf accretion-induced collapse, supernova core collapse,
gamma ray bursts, and the stochastic background. In this paper we
consider an additional possibility, GWs produced by unbound orbits.
As we will show, among unbound orbits near-parabolic encounters
(PEs) produce gravitational radiation with typical frequencies
appropriate for detection with terrestrial facilities. For close PEs
the gravitational radiation is short and intensive, that is
observable to large distances. Here, we estimate the expected event
rate of detections for specific current and near-future GW
detectors.

Initial order-of-magnitude estimates on the detectability of GWs
emitted during scattering and near collisions of stellar mass
compact objects in active galactic nuclei and globular clusters
(GCs) were made by \cite{dpz82}. Although their study primarily
focused on BH--star, star--star encounters, and did not provide
numbers for BH--BH encounters, they identified these encounters to
be ``quite rare''. However, \cite{dpz82} used an overly simplified
GC model in which the velocities and masses of all objects were
identical, and the spatial distribution was assumed to be
homogeneous. We extend the detection rate estimates to account for
the stellar BH mass function, mass segregation, and mass-dependent
relative velocities. We show that this improvement significantly
increases the event rate, by approximately a factor of $10^2$. In
addition, interferometric GW detector technology has improved
greatly and detailed sensitivity curves are now available.
\cite{dpz82} estimated a maximum visible distance of
$D_{\max}=20\Mpc$, which is a factor of $\sim 100$ less than
Advanced LIGO's (AdLIGO) capabilities (see Fig.~\ref{fig:Dmax}
below), i.e. a factor of $\sim 10^6$ less in the accessible volume
of sources. Combining these factors, our detection rate estimates
yield $\sim10^{8}$ times larger results for AdLIGO.

Gravity waves emitted during PEs are also important for creating
relativistic orbits by gravitational radiation reaction around the
supermassive black holes (SMBH) in the centers of galaxies. The GWs
emitted by the later inspiral of the star or compact object around
the SMBH are possibly detectable by the future space detector {\it
LISA} \citep{sr97,frei03,gair04} and also by ground-based detectors
for highly eccentric orbits \citep{ha05}. In the present paper, we
do not consider encounters with SMBHs, but focus on the direct
detection of GWs from unbound orbits of two stellar mass compact
objects (COs).

Stellar mass unbound orbit encounters are expected to be most likely
from dense star clusters with a large fraction of COs. Among regular
star systems, these features are carried by galactic nuclei and GCs,
where central densities reach $10^{4}$ -- $10^{7}\pc^{-3}$ within a
region of $0.5$ -- $3\pc$ \citep{pm93}, the inner regions contain a
CO fraction of $q\gsim 1/2$ \citep{sp95}. In the present paper, we
estimate PE event rates for GCs.

As compared to other GW burst sources, the big advantage in
detecting PE events is that the possible signal waveforms are much
more reliable as the physics behind them is well understood. The
waveforms are known analytically for the case of arbitrary masses
moving with arbitrary velocities but at small deflection angles
(often referred to as gravitational bremsstrahlung, see
\citealt{kt78}), arbitrary unbound orbits but low velocities in the
Newtonian approximation \citep{Turner}, in the post-Newtonian
approximation \citep[PN,][including corrections ${\cal
O}(v^2)$]{bs89}, in the 2PN approximation \citep[${\cal
O}(v^4)$][]{bdiww95,mvg05}, and most recently, in the 3PN
approximation \citep[${\cal O}(v^6)$]{bdei05}, and the exact
numerical solution is available for extreme mass ratios using a
Schwarzschild background approximation \citep{mar04}, and finally,
for head-on collisions with large velocities \citep{dp92}. Thus, PE
waveforms are available for a very large portion of the parameter
space. Waveform templates can be constructed a priori, similar to
inspirals. The prior knowledge of the possible waveforms allows the
method of matched filtering detection, which helps to reduce the
minimum signal-to-noise ratio necessary for a confirmed detection
\citep{fh98}.

This paper is organized as follows.
In \S~\ref{sec:Detectors}, we summarize the relevant characteristics
of interferometric GW detectors.
In \S~\ref{sec:PE waveforms}, we review the PE waveforms that we
adopt.
In \S~\ref{sec:Models}, we describe the population models that are
necessary to estimate the scalings of parameters and the overall PE
event rates.
In \S~\ref{sec:event rate}, we derive the expected number of PE
event rates, calculate their maximum distance of detection, and
estimate the implied rates of successful detections.
Finally, in \S~\ref{sec:Conclusions} we summarize our conclusions
and in \S~\ref{sec:Discussion} discuss the limitations and the
implications of this work.

\section{Overview of Gravitational-wave Detectors}\label{sec:Detectors}

The new generation of GW detectors rely on interferometric
monitoring of the relative (differential) separation of mirrors,
which play the role of test masses, responding to space-time
distortions induced by the GWs as they traverse the detectors. The
effect of a GW is to produce a strain in space, which displaces the
mirrors at the ends of the arms by an amount proportional to the
effective arm length and GW strain. For GWs incident normal to the
plane of the detector, and polarized along the arms of the detector,
the mirrors at the ends of the two arms experience differential
motion. Waves incident from other directions and/or polarizations
also induce differential motion, albeit at a smaller level.

Presently, there is an operational international network of first
generation interferometric GW detectors: InLIGO, VIRGO, TAMA, and
GEO (see \S~\ref{sec:intro} for references). The design of advanced
terrestrial GW detector AdLIGO and space detector {\it LISA} is well
on the way. There are also plans for a new generation of
low-frequency underground detectors especially sensitive for lower
frequencies \citep{desalvo04}, which might be especially sensitive
to PEs, which we will discuss in a separate paper in detail.
Finally, there are plans for possible future improvements of space
detectors: {\it Decihertz Interferometric Gravitational-wave
Observatory (DECIGO)} \citep{DECIGO}, {\it Advanced Laser
Interferometer Antenna (ALIA)} and the {\it Big Bang Observer (BBO)}
\citep{cc05b}. Their sensitivities, detection frequency bands and
capabilities are quite different. For our purposes, a good
approximation is to use: (1.) the InLIGO and VIRGO sensitivity goal
(nearly reached) to assess present capabilities; (2.) the AdLIGO
sensitivity goal to assess future capabilities of ground based
detectors; (3.) the {\it LISA} sensitivity goal to assess future
capabilities of initial space based detectors; and (4.) the {\it
Next Generation LISA} sensitivity goal to assess the capabilities of
possible further extensions to space detectors.
\begin{figure}
  \centering\mbox{\includegraphics[width=8cm]{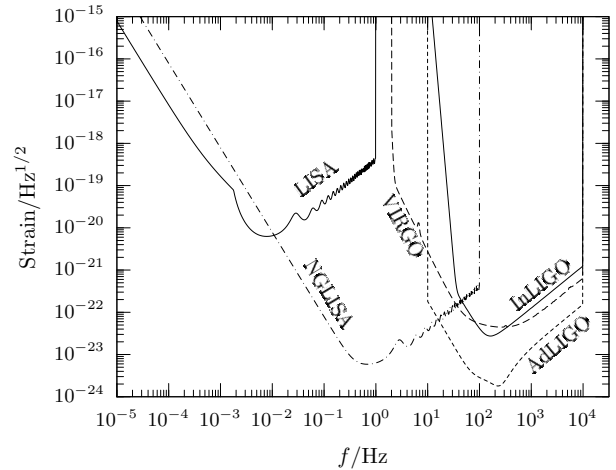}}
  \caption{\label{fig:sensitivity} Goal sensitivity curves for
  interferometric GW detector facilities: InLIGO, VIRGO, AdLIGO,
  {\it LISA}, and {\it NGLISA}.}
\end{figure}


The goal RMS noise density per frequency interval for the various
detectors, including instrumental and confusion noise, is plotted on
Figure~\ref{fig:sensitivity}.  For LIGO we adopt \cite{a04,lw96},
for VIRGO we adopt \cite{virgo05} but for simplicity discard the
narrow features\footnote{${\rm
http://www.virgo.infn.it/senscurve/}$}, for AdLIGO we adopt the
noise estimates from its website\footnote{${\rm
http://www.ligo.caltech.edu/advLIGO/scripts/ref\_des.shtml}$}, and
for {\it LISA}, we utilize the online sensitivity curve generator
\footnote{${\rm http://www.srl.caltech.edu/\sim
shane/sensitivity/}$} for the instrumental noise, and adopt the
confusion noise estimate from \cite{bc04a}. The noise levels of
possible extensions to {\it LISA} named the ``Next Generation {\it
LISA}'' ({\it NGLISA}) are also provided by the sensitivity curve
generator, which we also include in all of our calculations. This is
very similar to the planned sensitivity curve of {\it DECIGO}
\citep[see the more conservative case therein]{DECIGO} and is just
halfway between  {\it ALIA} and {\it BBO} \citep[a factor of 3
difference in sensitivity from both]{cc05b}.

\section{Parabolic Encounter Waveforms}
\label{sec:PE waveforms}

The GW signal waveform for PE is available in a wide range of
approximations (see \S~\ref{sec:intro}). We adopt \cite{Turner} for
the angular averaged waveforms, for which the interacting masses
travel on classical Newtonian trajectories and emit quadrupole
radiation. Other features such as spin-spin, spin-orbit
interactions, and gravitational recoil, etc., are higher order
perturbations which carry only a small total signal power in typical
cases. Therefore for the sake of calculating the signal-to-noise
ratio, it is a sufficient first-order approximation to use these
waveforms.

Illustrative examples of PE waveforms can be found in \cite{Turner}
(see Fig.~4 and 7 therein). The waveforms are generally constituted
of a large amplitude single peak or a jump in the time domain with
characteristic time scale $t_0$, related to the relative angular
velocity at the minimum separation $\omega_0=v_0/b_0=1/t_0$. Here
$v_0$ is the relative velocity at the closest point, and $b_0$ is
the corresponding minimum separation. \cite{Turner} provides a
closed analytical formula for the total GW radiation energy spectrum
$\D E/\D f$. The spectrum is wide-band, for parabolic orbits it is
zero at $f=0$, it has a maximum near $f_0=\omega_0/2\pi$ and a
half-width $\sim 1.5 f_0$.

The characteristic signal amplitude is obtained from the GW energy
spectrum as \citep{thorne87,fh98}
\begin{equation}\label{eq:h(f)0}
h(f)=\frac{\sqrt{3}}{2\pi}\frac{\G^{1/2}}{\cl^{3/2}}\frac{1+z}{\dL(z)}
\frac{1}{f}\sqrt{\frac{\D E}{\D f}[(1+z)f]},
\end{equation}
where $z$ is the redshift, $\dL(z)$ is the luminosity distance
\citep{eis97}, and $\D E/\D f\;[(1+z)f]$ is the total GW emitted
energy of the source at the emitted frequency. The orientation
averaged signal-to-noise ratio is
\begin{equation}\label{eq:snr}
\left\langle\frac{S}{N}\right\rangle =
\sqrt{\frac{4}{5}\int_{0}^{\infty} \frac{|h(f)|^2}{\hn(f)^2}\,\D f }
\end{equation}
where $h(f)$ is the characteristic signal amplitude (\ref{eq:h(f)0})
and $\hn(f)$ is the one-sided spectral noise density (see
\S~\ref{sec:Detectors} for references for the particular detectors).
Note that the relation (\ref{eq:snr}) refers to an angle-averaged
SNR obtained from the cube-root of an average of cubed signal
amplitudes over different possible orientations of the source and
interferometer. Since event rates roughly scale with volume, i.e.
distance cubed or $(S/N)^{-3}$, this prescription is useful for
estimating event rates \citep{thorne87}. For signals with optimal
orientations, the coefficient $4/5$ in (\ref{eq:snr}) is changed to
4. Note furthermore, that the $4/5$ factor is applicable for the
detection rate using a single interferometric GW detector. There are
already 4 interferometric GW detectors on Earth (see
\S~\ref{sec:Detectors}), and it is possible that there will be a lot
more in the future. A coincident analysis with multiple detectors
can be used to improve the efficiency by increasing the total
signal-to-noise ratio and also by insuring that at least one
detector is close to the optimal orientation \citep{jkkt96}. For
this reason, the coefficient $4/5$ in (\ref{eq:snr}) is most likely
pessimistic. For 1 detector in the optimal orientation and $K-1$
identical detectors in random orientations a quick scaling of the
coefficient is $\sim 4 + (4/5) (K-1)^{1/2}$. On the other hand, a
relatively large $\rm SNR$ might be required to keep the false alarm
rate at a sufficiently low level. For a conservative estimate on the
PE rate we do not modify the $4/5$ factor in the definition of $\rm
SNR$ (\ref{eq:snr}) and evaluate results for ${\rm SNR}=5$.

The PE waveforms can be obtained from eq.~(\ref{eq:h(f)0}) by
substituting the $E(f)$ relationship specific for PEs using
\cite{Turner}:
\begin{equation}\label{eq:h(f)1}
h(f,f_0)= \frac{\sqrt{85}\,\pi^{2/3}}{2^{5/3}}\frac{\G^{5/3}}{\cl^4}
\frac{{\Mchirp_z}^{5/3}}{\dL}\frac{f_{0z}^{2/3}}{f}\sqrt{F(f/f_{0z})}
\end{equation}
where $\Mchirp_z=(1+z) (m_1m_2)^{3/5}/(m_1+m_2)^{1/5}$ is the
redshifted chirp mass if $m_1$ and $m_2$ are the masses of the
interacting objects, $f_{0z}=f_0/(1+z)$ is the redshifted
characteristic frequency, f denotes the observed GW frequency, and
$F(x)$ is the \cite{Turner} normalized dimensionless energy spectrum
for dimensionless frequency $x=f/f_{0z}$, for which
$\int_0^{\infty}F(x)\,\D x=1$.

Equation~(\ref{eq:h(f)1}) is the leading order (i.e. Newtonian)
approximation to the GW waveform, $h(f,f_0)$. A remarkably
advantageous feature of the waveform in this approximation is that
it depends on only a single combination of the orbital parameters,
$f_{0z}$. Although we need only utilize this form (\ref{eq:h(f)1}),
we briefly note that it is also possible to express the waveform
with the separation at closest point $b_0$ in the center of mass
frame \citep{Turner}:
\begin{equation}\label{eq:h(t)2}
h(f,b_0)= \frac{\sqrt{85}}{4}\frac{\G^2}{\cl^4}\frac{(1+z)\,m_1
m_2}{\dL\, b_0 }\frac{1}{f}\sqrt{F[f/f_{0z}(b_0)]}.
\end{equation}
Here $f_{0z}(b_0)$ is the redshifted characteristic frequency for
fixed initial velocity as a function of $b_0$. The GW amplitude
spectrum $h(f,b_0)$ is roughly flat at low frequencies $f<f_{0z}$,
and decreases for higher frequencies. Equation~(\ref{eq:h(t)2})
shows that $h(f,b_0)$ scales with $b_0^{-1}$ for frequencies larger
than the cutoff at $\sim f_{0z}$.

Modifications are necessary for relativistic encounters. The
relativistic gravitational radiation waveforms and energy output has
been calculated by \cite{mar04} in the quadrupole approximation for
a test particle approaching a Schwarzschild black hole from infinity
on a quasi-parabolic geodesic\footnote{We continue to denote as
``quasi-parabolic'' or simply ``parabolic'' encounters that have
asymptotically zero velocity at infinity. Note that the trajectories
are generally quite different from parabolas in the highly
relativistic regime \citep[for illustration, see][]{mar04}.}. In
case the periastron distance $b_0$ is close to the unstable circular
orbit, the GW energy is significantly increased (it has a
logarithmic singularity at $\lambda=2$, where $\lambda\equiv
b_0/R_{\rm SH}$ and $R_{\rm SH}$ is the total Schwarzschild radius,
and a factor of $\sim 10$ increase for $\lambda = 2.01$ or a factor
of $2$ for $\lambda=3$). We adopt the fitting formula of
\cite{gkl05a}, which is correct within $0.1\%$ for orbits that avoid
a collision, and scale the amplitudes of the \cite{Turner} waveforms
(\ref{eq:h(f)1}) according to the increase of the GW energy, $E_{\rm
rel}(\lambda)/E_{\rm nr}(\lambda)$.
\begin{equation}\label{eq:h(f)2}
h(f,f_0)= \frac{\sqrt{85}\,\pi^{2/3}}{2^{5/3}}\frac{\G^{5/3}}{\cl^4}
\frac{{\Mchirp_z}^{5/3}}{\dL}\frac{f_{0z}^{2/3}}{f}
\sqrt{F\left(\frac{f}{f_{0z}}\right)} \sqrt{\frac{E_{\rm
rel}(\lambda)}{E_{\rm nr}(\lambda)}}
\end{equation}
Note once again, that the dimensionless periastron distance,
$\lambda$, is uniquely specified by the characteristic frequency,
$f_0$. We derive explicit formulae for $f_0(\lambda)$ in the
non-relativistic Newtonian and relativistic geodesic approximations,
when considering the dynamics of PEs in \S~\ref{sec:sub:rate1} and
\S~\ref{sec:sub:relativistic} below. Quite remarkably, the
characteristic frequency of the waveform is unchanged for
relativistic orbits. Therefore, we do not explore the effects of
relativistic modifications in the shape of the GW signal waveform,
we restrict only to correcting the amplitude.

\section{Population Models}
\label{sec:Models}

The major contribution to the PE event rate is expected from dense
star clusters with a large CO fraction. The most important systems
carrying these properties are possibly GCs and galactic nuclei. In
this paper we focus on GCs, but simple analytical scaling of the
results allows a straightforward extension to galactic nuclei or any
other population of spherical star systems.

The spatial distribution of GCs exactly traces the distribution of
galactic halos \citep{cwl04} in the local universe. In this section
we summarize the galaxy distribution and the GC abundance per
galaxy, and describe the GC models which we adopt for the paper.

\subsection{Galaxy Distribution}
We utilize the local distribution of galaxies \citep{Tul88}. The
accumulated number of galaxies to distance $D$ can be well
approximated by
\begin{equation}\label{eq:N_gal}
N^{\gal}(D)=\left\{\begin{array}{ll}
  N_1\; (D/\Mpc)^{0.9}   & \mbox{for~~} D \leq 3 \Mpc \\
  N_2\; (D/3\Mpc)^{1.5}  & \mbox{for~~} 3 \Mpc < D < 16\Mpc \\
  N_3\; (D/16\Mpc)^{2.4} & \mbox{for~~} 16 \Mpc < D < 60\Mpc \\
  N_4\; (D/60\Mpc)^3     & \mbox{for~~} D > 60 \Mpc\\
\end{array}
\right.,
\end{equation}
where $N_1$, $N_2$, $N_3$, and $N_4$ are 23, 62, 1100, and 26000,
respectively. The average density of faraway galaxies is
$0.03\Mpc^{-3}$, but the local galaxy abundance is much denser than
average. In Eq.~(\ref{eq:N_gal}) $N^{\rm gal}(D)$ has a $45\%$ jump
at the Virgo cluster at $D=16\Mpc$.

\subsection{Globular Cluster Abundance}
\label{sec:sub:GCabundance}

Following \cite{pzm00}, we adopt $\bar{n}^{\gc}=2.9 \Mpc^{-3}$ for
the average GC abundance in the universe. We roughly account for the
clustering of GCs in the local universe by assuming that the
distribution of GCs follow the abundance of galaxies. This
assumption is consistent with observations \citep{gsbkme03,cwl04}
suggesting that the population of GCs represent a universal, old
halo population that is present around all galaxies. The number of
GCs within a distance $D$ is then
\begin{equation}\label{eq:N_gc}
N^{\gc}(D) = y^{\gc} N^{\gal}(D),
\end{equation}
where $y^{\gc}$ is a scaling constant relating the abundance of GCs
and galaxies. Using the large scale average $\bar{n}^{\gc}=2.9
\Mpc^{-3}$ \citep{pzm00} and $\bar{n}^{\gal}=0.029 \Mpc^{-3}$
eq.~(\ref{eq:N_gal}) for $D>60\Mpc$ we get $y^{\gc}=100$.
Alternatively, $y^{\gc}$ can be interpreted as the number of GCs per
galaxy averaged over all morphological types. Concerning PE
detection rates, we shall show in \S~\ref{sec:sub:dist} that typical
observation distances for terrestrial detectors are larger than the
clustering scale (\ref{eq:N_gal}). Therefore, the results are
sensitive to mainly the average abundance and are only slightly
increased by the local clustering of GCs.

The value $\bar{n}^{\gc}=2.9 \Mpc^{-3}$ for the average abundance is
a conservative assumption. In their quick estimate \cite{pzm00}
derived this value by adding up the contribution of galaxies of
morphological types Sab, E-S0, and blue ellipticals. Recently, 12
nearby edge on spiral galaxies were examined, resulting in much
larger numbers, reaching $\sim 1000-1300$ GCs for these particular
galaxies \citep{gsbkme03,cwl04}. In addition to the morphological
types considered in \cite{pzm00}, dwarf elliptical (dE) galaxies
also contribute to the overall GC numbers \citep{bergh05}. The GC
content of 69 dwarf elliptical (dE) galaxies have been estimated to
host about a dozen GCs per dE galaxy \citep{lmf04}. Therefore, our
results on detection rates correspond only to lower limits, which
has to be scaled linearly with $\bar{n}^{\gc}$ when more detailed
estimates become available.

The GC distribution given by (\ref{eq:N_gal}--\ref{eq:N_gc}) is only
valid for sub-cosmological scales. Assuming that $D$ denotes the
luminosity distance, $\dL$, in eqs. (\ref{eq:N_gal}--\ref{eq:N_gc})
which is a direct observable using the GW amplitude, the change in
the cosmological volume element decreases the average density.
\begin{equation}\label{eq:N_gccos}
N^{\gc}(\dL) = 4\pi \int_0^{z(\dL)} \frac{\partial V}{\partial
z\,\partial \Omega} y^{\gc} \bar{n}^{\gal} \;\D z
\end{equation}
We adopt the cosmological volume element \citep{eis97} for a
$\Lambda$CDM cosmology with
$(\Omega_m,\Omega_\Lambda,h)=(0.3,0.7,0.7)$ consistent with recent
observations of the {\it Wilkinson Microwave Anisotropy Probe} and
the Sloan Digital Sky Survey \citep{teg04}. We find that the
uncorrected volume element $\dL^2 \D\Omega \D \dL$ is reduced by a
factor of 0.7 for $z=0.1$ ($\dL=0.5\Gpc$) and by a factor of 0.053
for $z=1$ ($\dL=7\Gpc$). Since GCs are believed to represent an old
halo population in galaxies, we do not account for additional
possible cosmological evolution of the comoving GC abundance.

\subsection{Globular Cluster Models}
\label{sec:sub:GCmodels}

Here we define the GC models that we use to obtain typical PE
parameters and event rates. First we define the common features that
are the same for both of our models. We assume a total of $N^{\rm
tot}$ stellar mass stars, $m_{\rm star}=\Msun$, spherically
distributed within a typical radius $R_{\gc}$. Also within the
cluster, is a CO population consisting of $N_{\CO}\equiv q N^{\rm
tot}$ objects with $q\ll 1$ that move in the background
gravitational potential of the stars. We define the ``typical CO
mass'' as $\langle m_{\CO}\rangle=10 \Msun$. We adopt typical values
of $N^{\rm tot}=10^6$, $R_{\gc}=1\pc$, $N_{\CO}=10^3$, with
$q=10^{-3}$, and $\langle m\rangle \simeq \Msun$
\citep{dm94,pzm00,mil02}.

We construct two different models for the distribution of mass,
spatial coordinate, and velocities of stars. Model I is a simple
plausible model to get the scaling of PE event rates on different
cluster parameters. Here we assume a homogeneous spherical
distribution, and the COs within the cluster have the same mass and
magnitudes of velocities. In Model II, we refine the assumptions to
account for the distributions of masses, mass segregation, and
relative velocity distributions.

By comparing Models I and II, we find that while Model I gives the
correct $R_{\gc}$, $N^{\rm tot}$, $N_{\CO}$, and $f_0$ scalings, it
underestimates the total event rate of a single GC by 2 orders of
magnitudes. The comparison of Models I and II is necessary (i) to
understand the origin of the large increase in PE event rates as
compared to \cite{dpz82}, and (ii) to understand the impact on the
BH mass function of GCs on PE event rates. It is therefore
emphasized that the GC model assumptions have a crucial importance
when determining the PE event rate. We believe that our Model II
includes all of the essential features of GCs for the assessment of
PE rates, and uncertainties are comprised of the uncertainties in
the model parameters rather than additional fundamental
processes\footnote{For example, details like bound binary
populations and interactions are not essential for PEs, and Spitzer
instability and core collapse can be accounted for by choosing our
model parameters accordingly (see \S~\ref{sec:sub:CoreCollapse} for
a discussion).}.

\subsubsection{Model I}

For the most simple model, the spatial distribution is assumed to be
uniform within a characteristic radius $R_{\gc}$, and it is assumed
that all COs have the same mass $m_{\CO}=10\Msun$ and magnitude of
velocity $v_{\vir}$. The orientation of velocities is isotropic,
implying a velocity dispersion of $\sigma \simeq v_{\vir}$. For this
simple model, we estimate the relative velocity distribution of COs
to be the same as the individual velocities $v_{\rel} \simeq
v_{\vir}$.
The characteristic velocities can be obtained from the virial
theorem for a uniform distribution of mass\footnote{There is a
similar result for a spherical star system of polytropic
distribution with a root mean square radius R. The only difference
is in the $3/5$ factor of Eq.~{\ref{eq:v_vir}}, which becomes $1/2$
in that case (e.g. \citealt{Saslaw}).}:
\begin{equation}\label{eq:v_vir}
v_{\vir}=\sqrt{\frac{3}{5}\frac{\G N^{\rm tot}\langle m
\rangle}{R_{\gc}}}
\end{equation}

The PE event rate is calculated as the rate of scattering on a fixed
target lattice, with incident velocity $v_{\vir}$.

\subsubsection{Model II}
\label{sec:sub:sub:Model2}

\begin{figure}
  \centering{\mbox{\includegraphics[width=8cm]{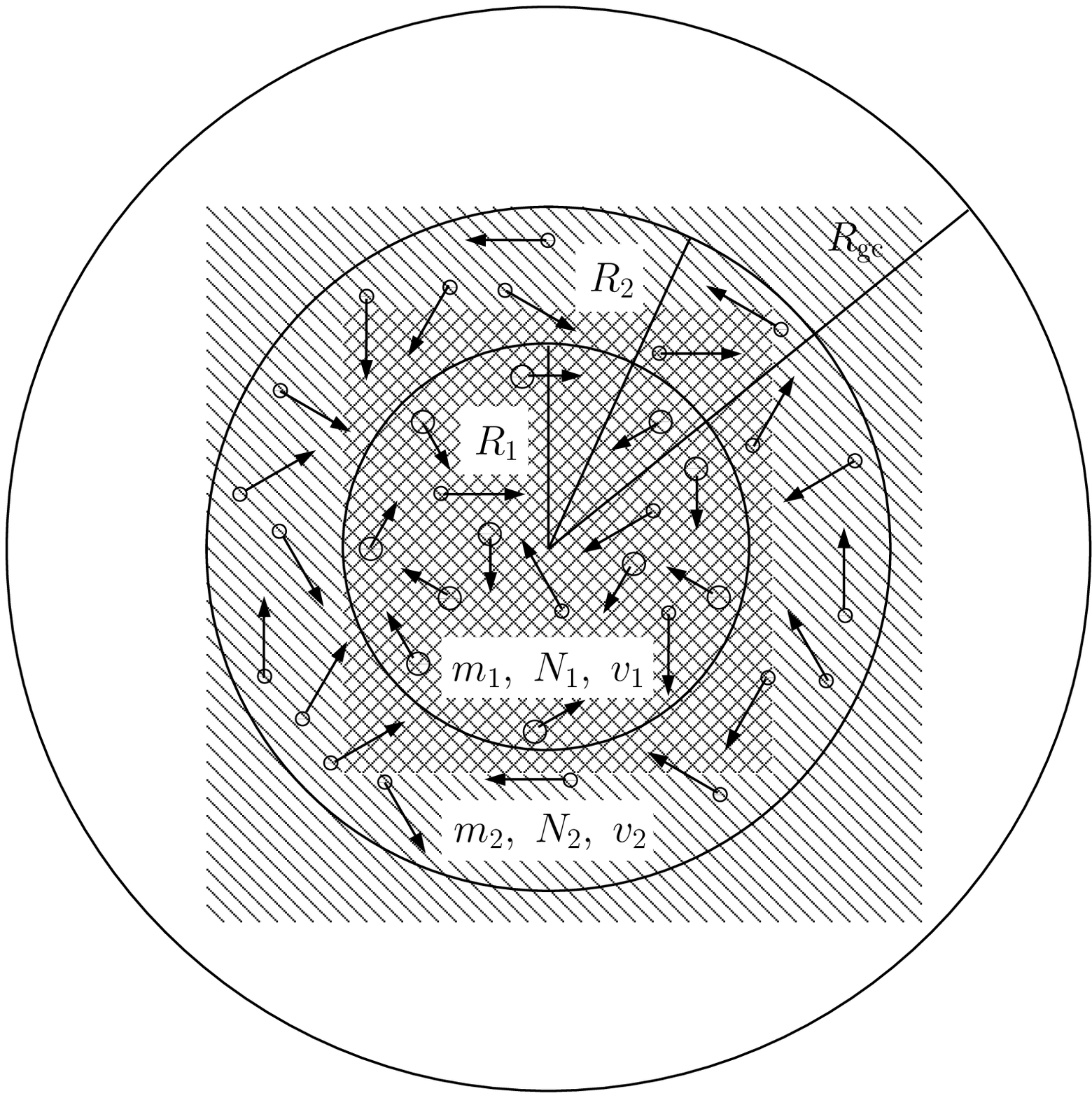}}}
  \caption{\label{fig:scattering} Encounters for Model II. More massive
  COs $m_1>m_2$ are distributed uniformly within a sphere of smaller
  radius $R_1<R_2$ and have smaller velocities. PEs between COs with
  $m_1$ and $m_2$ can take place within $R_1$. The relative velocity before the
  interaction is $v_\infty=v_{12}(m_1,m_2)$.
  Note that Model II assumes a continuous mass function, for which
  $N_1$ and $N_2$ are in fact infinitesimal.}
\end{figure}

We improve Model I with the following factors. The validity and
motivation of these assumptions is discussed below the list.
\begin{enumerate}
 \item We assume an equal number of NSs and BHs
 $N_{\NS}=N_{\BH}$. We introduce $g_{\BH}(m)=0.5 [\ln(m_{\max}/m_{\min})]^{-1}\,m^{-1}$
 for the fractional distribution of BH masses among COs with mass $m$ per
 mass interval $\D m$, in the range $m_{\min}<m<m_{\max}$, with
 $\int g_{\BH}(m)\D m = 0.5$. We define the NS mass distribution
 $g_{\NS}(m)$ as a Gaussian distribution with norm 0.5, mean $1.35
 \Msun$, and variance $0.1 \Msun$. Finally we define the CO
 distribution by $g_{\CO}(m)=g_{\BH}(m)+g_{\NS}(m)$, which has
 a unit norm. For definiteness, we take $N_{\BH}=500$, $m_{\min}=5\Msun$, and
 $m_{\max}=60\Msun$, implying that $\langle m_{\BH}\rangle\simeq 20
 \Msun$ and $\langle m_{\CO}\rangle\simeq 10 \Msun$, but also
 calculate detection rates for more general BH mass distributions.
 The PE detection rates are practically independent of the actual
 total number and distribution of NSs.
 \item We account for the mass segregation by assuming
 thermal equipartition among COs, objects with mass $m$ have a
 velocity $v_m= (m/\langle m\rangle)^{-1/2}v_{\vir}$,
 and are confined within a radius,
 $R_{m}=(m/\langle m\rangle)^{-1/2}R_{\gc}$, while
 regular stars are distributed uniformly within a sphere of
 radius $R_{\gc}$, as in Model I. Since $N\gg N_{\CO}$, the background
 gravitational potential is determined by regular stars.  See
 Fig.~\ref{fig:scattering} for an illustration. For core collapsed
 models a modified scaling is necessary (see text below).
 \item The relative velocity for COs with masses $m_1$ and $m_2$ is
 assumed to be $v_{\rel}\equiv v_{12} = [(m_1^{-1}+ m_2^{-1}) \langle m\rangle]^{1/2}v_{\vir}$.
\end{enumerate}
The BH mass distribution is crucial for the analysis,
since signal rates scale with $m^{19/3}$ (see \S~\ref{sec:event
rate} and the appendix below). Unfortunately, the analysis of BH
mass functions have not yet converged. Recent X-ray observations
display evidence for 20 galactic BHs with masses between $4 \lsim
m/\Msun \lsim 14$ \citep{cas05}, and $\sim 45$ ultra-luminous X-ray
sources are identified with intermediate-mass black holes (IMBHs)
with masses $m = 10^2$ -- $10^4\Msun$
\citep{pc04,mc04,grh05,blecha05}. Theoretical predictions from
two-dimensional simulations of stellar core collapse \citep{fk01}
lead to masses smaller than $20\Msun$ with very different
distributions depending on the assumptions (fraction of explosion
energy used to unbind the star, stellar winds, mass transfer after
helium ignition). Sophisticated simulations of the initial phase of
rapid star evolution assuming a lower metallicity for the progenitor
stars (weaker stellar winds) appropriate for GCs and including a
large fraction of binaries, collisions, and accretion leading to the
mass buildup of BHs imply an initial smooth decreasing distribution
of stellar-mass BHs with masses up to $\sim 60$ -- $100\Msun$
\citep{belcz05} depending on model assumptions and cluster
environments. Results are valid for timescales short compared to
later dynamical evolution of the cluster. However, the final fate of
the cluster remains highly uncertain. In small GCs, dynamical
interactions of binaries might eject a significant portion of the
stellar-mass BH population \citep{sh93,pzm00,O'Leary05}. Following
\cite{mil02} and \cite{w04} we adopt $g_{\BH}(m)\propto m^{-1}$
leaving the minimum and maximum masses free parameters. However,
most recent population synthesis simulations \citep{belcz05}
typically yield steeper BH mass functions. For this reason we felt
it important to compute results for other distributions
$g_{\BH}(m)\propto m^{-p}$ with $p=0$, 1, and 2, as well. Concerning
the other parameters, in our standard model we take $m_{\min}=5
\Msun$ and $m_{\max}=60 \Msun$, for which $\langle
m_{\BH}\rangle\simeq 20 \Msun$ and $\langle m_{\CO}\rangle\simeq 10
\Msun$, and assume $N_{\BH}=500$, but also calculate results for
other $N_{\BH}$, $m_{\min}$, and $m_{\max}$ values.

The NS mass distribution that we utilize is supported by
observations of 26 radio pulsars and 4 X-ray binaries \citep[and
references therein]{pp01}. The distribution is sharply peaked around
$1.35 \Msun$. Our assumption on the total number of NSs
($N_{\NS}=500$) is somewhat arbitrary, values could be higher or
lower depending on what fraction of NSs are ejected by kicks during
their formation. Our detection rate estimates can be scaled to the
appropriate value using $\nu_{\NS-\NS}\propto N_{\NS}^2$ and
$\nu_{\BH-\NS}\propto N_{\NS}N_{\BH}$ (see \S~\ref{sec:event rate}
below). Note however, that in \S~\ref{sec:event rate} we show that
NS interactions contribute a negligible fraction of the PE event
rate (see Figs.~\ref{fig:ratemass} and \ref{fig:ratemassratio}
below).

Note that we do not consider encounters between white dwarfs (WDs).
The detection of PEs between WDs are even worse than for NSs, since
WD masses are even smaller \citep{bt87}. Our final results show that
the PE detection rates roughly scale with $\sim m^{8.33}$, implying
that rates are enormously suppressed for WDs: by a factor of $\sim
10^{15}$ relative to $50\Msun$ BHs, which are typical for detection
(see Fig.~\ref{fig:ratemass}). Moreover, WDs are disrupted by tidal
torques for close encounters, at sub-Hz frequencies. Therefore, PEs
of WDs are completely invisible for terrestrial detectors.

The second improvement is to account for mass segregation. The
differentiation of the stellar population with mass within the
cluster core is a consequence of thermal equilibrium \citep{bt87}.
Objects with masses larger than $10\Msun$ had enough time to relax
to thermal equilibrium within the lifetime of the GCs \citep{fs82}.
This means that the kinetic energies of each of the component stars
are drawn from the same distribution, implying that the typical
speed of an object of mass $m$ is $v_m = (m/\langle
m\rangle)^{-1/2}v_{\vir}$, causing the object to sink to the core of
the cluster. For a nearly homogeneous distribution of background
stars, this implies that the maximum radius available to a given
mass is $R_m=(m/\langle m \rangle)^{-1/2}R_{\gc}$ \citep{bt87}. For
the BH distribution given by $g_\BH(m)$ we get that BHs with mass
$\sim 50\Msun$ (which make the dominant contribution to PE rates,
see Figs.~\ref{fig:ratemass} and \ref{fig:ratemassratio} below) are
confined to a radius $\sim 0.14R_{\gc}$. Note that our scalings
based on thermal equipartition might not hold in case Spitzer
instability leads to core collapse, creating a dynamically decoupled
core of high mass BHs (typically $R_{\rm core}\sim 0.01$--$0.10
R_{\gc}$ depending on the fraction of primordial binaries,
\citealt{hth06}). We can account for core collapse by simply scaling
our final results on detection rates appropriately with $N^2_{\BH}
R_{\rm core}^{-3}v_{\rm core}^{-1}$ (see
\S~\ref{sec:sub:CoreCollapse} for a discussion).

Finally, we discuss the assumption on the relative velocity. The
velocity distribution of stars in GCs is well described by the
King-Michie (KM) model \citep{mey87}, which is roughly a
Maxwell-Boltzmann (MB) distribution with a maximum velocity cutoff.
It is well-known that the relative velocity distribution for MB
individual velocity distributions is also MB for the reduced mass
$\mu=m_1m_2/(m_1+m_2)$ \citep{bt87}. Thus $\langle
v_{12}\rangle_{\rm RMS} = (m_1/\mu)^{1/2}\langle v_{m_1}\rangle_{\rm
RMS}=[(m_1^{-1}+ m_2^{-1}) \langle m\rangle]^{1/2}v_{\vir}$.

Note that we do not utilize the exact velocity distribution, but
associate the same fixed velocity value for every object with
identical masses. Relaxing this approximation and accounting for MB
velocity distributions leads to a change of only a few percent in
the encounter rate results (the correction is $(3/\pi)^{1/2}$ for
Model I in the range of GW detector frequencies, see \citealt{kg04}
for the derivation). Thus, the velocity distribution can be safely
approximated with the mean value. A simple explanation is the fact
that GW detector frequencies correspond to unbound encounters with
nearly parabolic trajectories, for which the exact value of the
initial velocity is negligible (see \S~\ref{sec:sub:rate1} below).

The PE event rate for component masses $m_1$ and $m_2$ is calculated
as the rate of scattering with incident velocity equal to the
initial relative velocity $v_{\rel}\equiv v_{12}(m_1,m_2)$.

\section{Parabolic Encounter Event Rate}
\label{sec:event rate}

We now derive the event rate for the successful detection of PE
signals using the two models of GCs. This section is divided in five
parts.
In \S~\ref{sec:sub:rate1} we derive the comoving event rate per
comoving characteristic frequency bins for individual GCs for the
two population models.
In \S~\ref{sec:sub:relativistic} we derive the modifications
necessary for relativistic encounters.
In \S~\ref{sec:sub:dist} we determine the signal-to-noise ratio
using the specific detector sensitivity curves and determine the
maximum observable distance of PEs.
In \S~\ref{sec:sub:rate3}, we add up the contributions of all
possible GCs within the visible distance and estimate the PE
detection rates.
Finally in \S~\ref{sec:sub:rateresult} we conclude the results of
the analysis.

\subsection{Contributions of Individual Globular Clusters}
\label{sec:sub:rate1}

\subsubsection{Outline}
When calculating the detectable event rates for specific GW
detectors, it is desirable to express the encounter cross section
for particular $f_0$ frequency bins. To achieve this we first
compute the interaction cross-sections for given masses and initial
orbital parameters. Then, the Newtonian equations of motion relate
the initial orbital parameters to $f_0$. Changing to the $f_0$
variable leads to the partial event rate for the given masses and
characteristic encounter frequencies. Then, for Model I, it is very
simple to add up the individual contributions of all objects within
the cluster. For Model II, we utilize the specific radial and
relative velocity distributions, $R_m$ and $v_{12}$, and average
over the CO mass distribution.

\subsubsection{Derivation of Event Rates}\label{sec:sub:sub:DerivationGC}
The typical minimum distance between COs in these systems is
$R_{\gc}/\sqrt[3]{N}\simeq 10^{11}\km$, a value several orders of
magnitudes larger than the typical minimum separation of an
encounter ($10\km < b_0 \lsim 10^7\km$ for detectable frequencies,
see Eq.~[\ref{eq:b0}] below). Therefore, a sufficient approximation
is to consider short-time two-body interactions during
encounters\footnote{We shall discuss below that the PE event rate is
not sensitive on whether the interacting participants are elements
of regular bound binaries or if they are single objects.}, and
constant velocities in between events. The PE event rate can then be
simply estimated by a scattering of particles with incident initial
velocities $v_{\infty}=v_{\rel}$ on a still target lattice.

Since the velocities are assumed to be locally isotropic everywhere
in the cluster, the cross section of a particle with an impact
parameter between $b_{\infty}$ and $b_{\infty} + \D b_{\infty}$ is
$\D\sigma=2\pi b_{\infty}\D b_{\infty}$. We proceed to express the
infinitesimal cross section for $\D f_0$ bins.

We derive encounter parameters with a non-relativistic Newtonian
description. The separation, $b_0$, and relative velocity, $v_0$, at
periastron can be computed from the initial conditions of the
interacting bodies. The initial parameters are the impact parameter
$b_{\infty}$ and the velocity, $v_{\infty}\equiv v_{\rel}\equiv
v_{12}(m_1,m_2)$, of the scattered particle (see
\S~\ref{sec:sub:sub:Model2} for the definition). Using the
conservation of mechanical energy and angular momentum we get
\begin{eqnarray}\label{eq:parameters@infty}
    b_{\infty} =  \frac{b_0}{\sqrt{1-2\gamma}},&\quad\mbox{and}\quad&   v_{\infty} = v_0
    \sqrt{1-2\gamma},
\end{eqnarray}
where $\gamma=(\G M)/(b_0{v_0}^2)$ is the ratio of potential energy
and double kinetic energy at the closest point\footnote{$\gamma<1/2$
for hyperbolic Newtonian trajectories.}, with $M=m_1+m_2$. Although
Eq.~(\ref{eq:parameters@infty}) is strictly only valid in the
comoving reference frame of the center-of-mass, it is an adequate
approximation for the realistic parameters as long as $v_\infty\ll
v_0$. Again, we point out that the relevant encounters are nearly
parabolic, so that the initial velocity distributions have a
negligible impact on the result.

Let us express the cross section with variables $b_0$ and $v_\infty$
using Eq.~(\ref{eq:parameters@infty}) and make use of our
simplifying assumption that $v_{\infty}$ is a constant for fixed
masses (see \S~\ref{sec:sub:GCmodels}). The result is
\begin{equation}\label{eq:cross section}
\D\sigma = \left(\frac{\G M}{v_{\infty}^2 b_0} + 1\right)2\pi b_0 \D
b_0.
\end{equation}
Here, the first term dominates the parenthesis for typical
$v_{\infty}$ and $b_0$ values (see Eq.~\ref{eq:b0}. below). This
term is responsible for the deflection of trajectories due to
gravity.

The characteristic GW frequency is directly related to the minimum
separation, $b_0$, and the relative velocity $v_0$ at $b_0$ by
\begin{equation}\label{eq:fdef}
\omega_0=2\pi f_0= \frac{v_0}{b_0}.
\end{equation}
Using Eq.~(\ref{eq:parameters@infty}),
\begin{equation}\label{eq:f(b_0,v_1)}
    \omega_0 = (2 \G M)^{1/2}b_0^{-3/2}\left(
    1+\frac{v_{\infty}^2 b_0}{2 \G M}\right)^{1/2}.
\end{equation}

For typical frequency bands of InLIGO ({\it LISA}) the correction on
the RHS is $v_{\infty}^2b_0/(2\G M_{\CO})\lsim 10^{-8}$ ($10^{-4}$)
for InLIGO ({\it LISA}). In order to get the event rate per
frequency bin, we need the inverse relationship
$b_0(\omega_0,v_{\infty})$. This can be obtained recursively, as a
power series in $\omega_0$. To the second non-vanishing order, we
get
\begin{equation}\label{eq:b0}
b_0 = (2 \G M)^{1/3}\omega_0^{-2/3}
\left(1+\frac{1}{3}\frac{v_{\infty}^2}{(2 \G
M)^{2/3}}\omega_0^{-2/3}\right).
\end{equation}
Substituting in Eq.~(\ref{eq:cross section}) yields
\begin{equation}\label{eq:dsigma}
\D\sigma = \frac{2\pi}{3} \frac{(2 \G M)^{4/3}}{v_{\infty}^2}
\omega_0^{-2/3} \left(1 + \frac{8}{3} \frac{v_{\infty}^2}{(2 \G
M)^{2/3}}\omega_0^{-2/3}\right) \frac{\D \omega_0}{\omega_0}.
\end{equation}

For Model I, the scattering rate for a single particle is $n_{\CO}
v_{\infty}\D\sigma$, where $n_{\CO}$ is the number density of COs.
Since there are a total of $N_{\CO}$ particles, the contribution of
all COs is $1/2\times N_{\CO}n_{\CO} v_{\infty}\D\sigma$, where the
$1/2$ factor comes from the fact that the same ensemble of particles
constitute both the targets and the injection. Thus
Eq.~(\ref{eq:dsigma}) becomes
\begin{equation}\label{eq:dnu1}
\D\nu^{\I} = \nu^{\I}_1\; \left(\frac{f_0}{f_{100}}\right)^{-2/3}
\left(1 + \left(\frac{f_0}{f_1^{\I}}\right)^{-2/3}\right) \frac{\D
f_0}{f_0}.
\end{equation}
where $f_{100}=100\Hz$ and
\begin{eqnarray}\label{eq:nu1I}
\nu^{\I}_1&=&(2\pi)^{-2/3}\frac{N_{\CO}^2}{4}\frac{(4 \G
M_{\CO})^{4/3}}{R_{\gc}^3 v_{\infty}}f_{100}^{-2/3}=6.7\times 10^{-15}{\yr}^{-1}\\
\label{eq:f1I} f_1^{\I} &=&
\frac{1}{2\pi}\left(\frac{8}{3}\right)^{3/2} \frac{v_{\vir}^3}{4\G
M_{\CO}} = 1.7 \times 10^{-8}\Hz.
\end{eqnarray}

For Model II, the interacting objects with masses $m_1$ and $m_2$
are distributed uniformly within radii $R_1$ and $R_2$ and have
mass-dependent relative velocities $v_{\infty}=v_{12}$ (see
Fig.~\ref{fig:scattering}). Let $N_1$ and $N_2$ denote the number of
particles with mass $m_1$ and $m_2$, respectively. Let us assume
$m_1>m_2$, for which $R_1<R_2$. In this case the interactions
between masses $m_1$ and $m_2$ take place only within a radius
$R_1$, where the density of particles with mass $m_2$ is
$n_2=N_2/(R_2^3 4\pi/3)$. For a smooth distribution, $N_1$ is the
infinitesimal number of particles with masses between $m_1$ and
$m_1+\D m_1$, i.e. $N_1=N_{\CO} g_{\CO}(m_1)\D m_1$ (and $N_2$
defined similarly). The scattering rate for an injection of $N_1$
particles with $v_{\infty}$ velocities on a target density $n_2$ is
is $N_1 n_2 v_{\infty}\D\sigma$. To get the total event rate for the
cluster for $\omega_0$ bins we need to integrate over the mass
distributions
\begin{align}\nonumber
\D\nu^{\II} &= \int_0^{\infty}\D m_1 g_{\CO}(m_1)
\int_0^{\infty}\D m_2 g_{\CO}(m_2)\;\times\\
&\,  \left[
\nu^{\II}_1(m_1,m_2)\left(\frac{f_0}{f_{100}}\right)^{-2/3} +
\nu^{\II}_2(m_1,m_2)\left(\frac{f_0}{f_{100}}\right)^{-4/3} \right]
\frac{\D f_0}{f_0},\label{eq:dnu2}
\end{align}
where $\nu^{\II}_1(m_1,m_2)$ and $\nu^{\II}_2(m_1,m_2)$ are given by
\begin{eqnarray}\label{eq:nu1II(m1,m2)}
\nu^{\II}_1(m_1,m_2) &=& (2\pi)^{-2/3}\frac{N_{\CO}^2}{4}
 \frac{(2 \G M)^{4/3}}{R_{>}^3 v_{\infty}}f_{100}^{-2/3}\\
\nu^{\II}_2(m_1,m_2) &=& (2\pi)^{-4/3}\frac{2N_{\CO}^2}{3}
 \frac{(2 \G M)^{2/3}v_{\infty}}{R_{>}^3}f_{100}^{-4/3}
\end{eqnarray}
where $R_> = \max(R_1,R_2)$ and the mass dependence is implicit in
the total mass $M$, $v_\infty=v_{12}$, $R_1$, and $R_2$ (see the
Appendix for explicit formulae).

The mass integrals in (\ref{eq:dnu2}) can be evaluated independent
of the frequency, resulting in the same functional form as for Model
I (\ref{eq:dnu1}). The constants for Model II are
\begin{eqnarray}\label{eq:nu1II}
  \nu_1^{\II} &=& 1.9 \times 10^{-12}\yr^{-1}\\\label{eq:f1II}
  f_1^{\II} &=& 1.0 \times 10^{-10}\Hz.
\end{eqnarray}
(see the Appendix for parametric formulae).

For Model II, it is also interesting to get the relative encounter
rates for $\BH-\BH$, $\BH-\NS$, and $\NS-\NS$ interactions.
Integrating Eq.~(\ref{eq:dnu2}) over the corresponding mass
intervals, we get
\begin{eqnarray}
 \nu^{\II}_{1,\BH-\BH} &=& 0.996 \; \nu^{\II}_{1} = 286 \; \nu^{\I}_{1}, \label{eq:nuBH-BH}\\
 \nu^{\II}_{1,\BH-\NS} &=& 4.14\times 10^{-3}\nu^{\II}_{1} = 1.19 \; \nu^{\I}_1,\label{eq:nuBH-NH}\\
 \nu^{\II}_{1,\NS-\NS} &=& 7.04\times 10^{-5}\nu^{\II}_{1} =0.02\;\nu^{\I}_1.\label{eq:nuNS-NS}
\end{eqnarray}
The corresponding analytical formulas are given in the Appendix. It
is clear that $\BH-\BH$ encounters dominate the event rates. The
event rates of $\NS-\NS$ encounters are more than four orders of
magnitude lower!

\subsubsection{Discussion}
Therefore we conclude that Model II has a much larger event rate
than Model I.  By inspection of Eq.~(\ref{eq:dnu2}) and
(\ref{eq:nu1II(m1,m2)}) the main factors responsible for this
increase can be identified. First, the CO density is increased by
the CO confinement in the core: $n\propto {R_m}^{-3}\propto
m^{3/2}$. Second, the typical CO relative velocity inverse is
increased: $v_{\infty}^{-1}\propto m^{1/2}$. Third, the
gravitational focusing is proportional to $m^{4/3}$. Thus,
$\nu_1\propto \left( R_{\gc}^3/R_{\CO}^{3}\right)\times
\left(\langle m^{10/3}\rangle/m_{\CO}^{10/3}\right)$. The PE event
rate is thus highly inclined towards the high-mass end of the BH
distribution in the cluster. For a more precise treatment, the exact
contributions of the component mass parameters are given in the
Appendix, Eqs.~(\ref{eq:BH-BH}-\ref{eq:NS-NS}). Note, that
\cite{dpz82} obtained results equivalent to our Eq.~(\ref{eq:nu1I})
of Model I, but they focus on star -- star encounters and use
$m=4\Msun$ instead of $m_{\CO}=10\Msun$. (Another difference is that
they do not discuss $f_0$-dependent differential rates, but derive
the total PE rate based on a typical minimum separation. Using
$m=4\Msun$ in Model I, the results are a factor of $10^3$ lower than
the rates for Model II.)

Equations~(\ref{eq:dnu1}-\ref{eq:f1I}) and
(\ref{eq:nu1II}-\ref{eq:f1II}) give the resulting PE event rate per
GC of all Newtonian trajectories between point masses for the two GC
models considered. A significant shortcoming is that COs have finite
radii and collide for sufficiently small minimum separations.
Moreover the Newtonian approximation breaks down for large
velocities or strong gravitational fields. These effects are
considered in \S~\ref{sec:sub:relativistic} below.

Notice how small is the correction proportional to $f_0^{-4/3}$ in
Eq.~(\ref{eq:dnu1}) for GW detector frequencies $f\gg f_1^{\I,\II}$.
Recall that in Eqs.~(\ref{eq:f(b_0,v_1)}) and (\ref{eq:b0}) the
expansion coefficient is proportional to $v_{\infty}$. Hence the
leading order term is exact for $v_{\infty}=0$, thus it corresponds
to parabolic trajectories. This proves our conjecture that the
unbound orbit encounter rate is dominated by near-PEs.

In Eq.~(\ref{eq:dnu1}) the leading-order terms are proportional to
$1/v_{\vir}$. The result is slightly counterintuitive if one
identifies the star system with an ideal gas, since for ideal gases,
the rate of collisions is directly proportional to $v_{\infty}$. In
this perspective it seems reasonable to expect the encounter rate to
be a growing function of $v_{\infty}$ for fixed frequency bins. The
confusion arises from the fact that our GC models are using the
opposite limit. For star systems the typical velocities are so small
that the gravitational interaction dominates the motion of the
stars.\footnote{The ideal gas model is sufficient only for extremely
small characteristic frequencies $f_0\ll f_1$
(eqs.~[\ref{eq:f1I}]~and~[\ref{eq:f1II}]), which is below the lower
frequency limit of GW detectors. In this regime the stars'
trajectories are only slightly deflected, implying that gravity, in
terms of encounter likelihood, is negligible.} Increasing the
velocities decreases the gravitational focusing, thereby decreasing
the encounter likelihood.

\begin{figure}
  \centerline{\includegraphics[width=8cm]{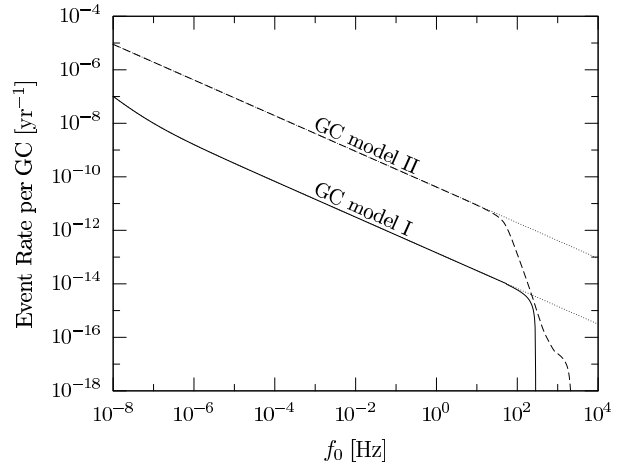}}
  \caption{\label{fig:latticerate} The expected total rate of PEs
  produced in a single GC per logarithmic frequency bin.
  GC Model I (solid) and Model II (dashed) results are shown including
  relativistic corrections for geodesics avoiding head-on collisions
  (see \S~\ref{sec:sub:relativistic}). The dotted lines represent
  PE event rates in the non-relativistic approximation. The
  non-relativistic treatment is adequate for low frequencies
  for which the trajectories avoid collisions with minimum separations of several
  Schwarzschild radii. Only a fraction of these events can be
  detected, depending on the distance of the GC.
  }
\end{figure}

The expected rate of PE events for a single GC is plotted in Figure
\ref{fig:latticerate} for logarithmic frequency bins for the two GC
models. The non-relativistic results presented in this section are
plotted with dotted lines, which overlap with the relativistic
calculation below $\sim 10\Hz$. For higher frequency the minimum
separation drops below $\sim 6$ Schwarzschild radii for the largest
BHs and relativistic corrections become important. Figure
\ref{fig:latticerate} displays that event rates are higher for lower
characteristic frequencies, e.g. for model II for
$f_0=0.1\,\mathrm{mHz}$ (the minimum frequency for space detectors),
we get $1.9\times 10^{-8}\yr^{-1}{\rm GC}^{-1}$ events, whereas for
terrestrial detectors $f_0=100\Hz$ it is only $1.9\times
10^{-12}\yr^{-1}{\rm GC}^{-1}$.

\subsection{Relativistic Orbits}\label{sec:sub:relativistic}

Up to this point the PE event rates have been estimated for fixed
characteristic frequencies but independent of the minimum separation
$b_0$ and relative velocity $v_0$. In addition to non-relativistic
parabolic encounters, these events also include head-on collisions,
relativistic captures, relativistic flybys, and zoom-whirl orbits.
Since we have used a Newtonian analysis in the derivation, our
results presented in \S~\ref{sec:sub:rate1} are valid for the
non-relativistic parabolic encounters only. Here, we improve the
classical calculation to account for general relativistic encounters
of test particles moving along geodesics in the Schwarzschild
space-time. This treatment is exact for extreme mass ratios, but as
an approximation we extrapolate these formulas for general
mass-ratios as well.

To classify the orbits, we introduce a parameter $\lambda\equiv
b_0/R_{\rm SH}$, where $b_0$ is the distance at periastron, and
$R_{\rm SH}=2 \G M/\cl^2$ is the Schwarzschild radius of the total
mass\footnote{Here we restrict to BH-BH encounters which dominate
event rates, see Eqs.~(\ref{eq:nuBH-BH}-\ref{eq:nuNS-NS}).}. For
Newtonian parabolic encounters, we get $\lambda = \cl^2/v_0^2$. We
distinguish (i) non-relativistic parabolic encounters for
$\lambda\geq 6$, (ii) general relativistic flybys for
$2.1<\lambda<6$, (iii) zoom-whirl orbits for $2\leq \lambda\leq
2.1$, and (iv) head-on collisions for $2<\lambda$. We restrict our
calculations to $\lambda\geq 2$, since this is the regime in which
matched filtering can be carried out using the waveforms of
\S~\ref{sec:PE waveforms}.

In this section, we improve \S~\ref{sec:sub:sub:DerivationGC} to
account for the relativistic deviations in the trajectories. In
practice, we repeat the derivation of
\S~\ref{sec:sub:sub:DerivationGC} to get the cross-section using the
orbital parameters of the geodesics of a test particle moving in a
Schwarzschild space-time \citep{gkl05a}.

For parabolic encounters the specific orbital angular momentum is
\begin{equation}\label{eq:L}
\widetilde{L}= \sqrt{2\G M b_0}\, \left(1-\lambda^{-1}
\right)^{-1/2},
\end{equation}
where $\lambda=\lambda(b_0)$ defined above. The non-relativistic
result is retained for $\lambda\rightarrow\infty$. Equating
(\ref{eq:L}) to the angular momentum before the encounter
$\widetilde{L}=b_{\infty}v_{\infty}$, solving for $b_{\infty}$, and
substituting in $\D\sigma=2\pi b_{\infty} \D b_{\infty}$ we get
\begin{equation}
\D\sigma = \frac{\G M}{v_{\infty}^2 b_0^2}
\frac{1-2\lambda^{-1}}{(1-\lambda^{-1})^2}\, 2\pi b_0 \D b_0.
\end{equation}
This is to be compared to the non-relativistic analogue
(\ref{eq:cross section}). The first term is the non-relativistic term
for near-parabolic orbits and the $\lambda$-dependent fraction
describes the relativistic correction. The latter decreases the
cross section per unit $b_0$. For $\lambda\rightarrow 2$ the cross
section becomes 0. For smaller impact parameters a head-on collision
takes place, for which the periastron distance and $\lambda$ is
undefined.

Repeating \S~\ref{sec:sub:sub:DerivationGC}, the next step is to
change to the $f_0$ characteristic frequency variable. Since
$\widetilde{L}=b_0^2\; \D \phi/\D \tau$, where $\D \tau =
(1-\lambda^{-1})^{1/2}\, \D t$ is the infinitesimal proper time
element along the geodesic at the closest approach
\citep[e.g.][]{Gravitation}, from (\ref{eq:L}) we get
\begin{equation}\label{eq:f0rel}
\frac{\D \phi}{\D t} \equiv \omega_0 \equiv 2\pi f_0= (2\G
M)^{1/2}b_0^{-3/2}.
\end{equation}
Quite remarkably, this is identical to the result of the
non-relativistic calculation for parabolic orbits
(\ref{eq:f(b_0,v_1)}). The GW waveforms have a peak at an angular
frequency $\omega_0$ for the non-relativistic encounters \citep[and
see \S~\ref{sec:PE waveforms} above]{Turner}. For relativistic
zoom-whirl orbits with several revolutions around the central BH,
the most intensive GWs are radiated at twice the orbital frequency.
It is also useful to get the inverse relationship from
eq.~(\ref{eq:f0rel}):
\begin{equation}\label{eq:lambda}
\lambda(M,f_0)= \left(\frac{\cl^3}{4\pi\G} \frac{1}{Mf_0
}\right)^{2/3}.
\end{equation}

According to Eq.~(\ref{eq:f0rel}), the non-relativistic result for
$\omega_0$ is adequate even in this regime. Using
(\ref{eq:f(b_0,v_1)}) we get
\begin{equation}\label{eq:dsigmarel}
\D\sigma = \frac{2\pi}{3} \frac{(2 \G M)^{4/3}}{v_{\infty}^2}
\omega_0^{-2/3}
\frac{1-(\omega_0/\omega_{M,\max})^{2/3}}{\left[1-\frac{1}{2}\,(\omega_0/\omega_{M,\max})^{2/3}\right]^2}
\frac{\D \omega_0}{\omega_0},
\end{equation}
analogous to (\ref{eq:dsigma}) for parabolic orbits, where
\begin{equation}\label{eq:fMmax}
\omega_{M,\max}\equiv 2\pi f_{M,\max} \equiv
\frac{2\pi\cl^3}{\sqrt{32}\G M}
\end{equation}
is the maximum angular frequency, corresponding to $\lambda=2$.

It is desirable to calculate the partial event rates of PEs with
minimum separations $b_0$ exceeding $\lambda R_{\rm SH}$, we
substitute in Eq.~(\ref{eq:f(b_0,v_1)}), and impose the resulting
constraint on the characteristic frequency:
\begin{equation}\label{eq:fconstraint}
f_0 \leq f_{M,\lambda}=2\pi\frac{\cl^3}{2\G M}\lambda^{-3/2}.
\end{equation}
For marginally plunging orbits, $\lambda= 2$, we get $f_{M,\lambda}=
f_{M,\max}$.

When adding up the total event rates for a particular $f_0$
(Eqs.~\ref{eq:dnu1} and \ref{eq:dnu2}) only the masses satisfying
the constraint (\ref{eq:fconstraint}) have to be included in the
mass integrals. Repeating \S~\ref{sec:sub:sub:DerivationGC} with
these modification, we get for Model I

\begin{equation}\label{eq:dnu1rel}
\D\nu^{I} = \nu^{I}_1 \left(\frac{f_0}{f_{100}}\right)^{-2/3}
\frac{1-2(f_0/f_{M,\max})^{2/3}}{\left[1-(f_0/f_{M,\max})^{2/3}\right]^2}
\frac{\D f_0}{f_0}.
\end{equation}
for $f_0 \leq f_{M_{\CO},\lambda}$, and $\D\nu^{I}=0$ otherwise. For
Model II, we get
\begin{align}\nonumber
\D\nu^{\II} &=& \int\hspace{-1ex}\int_{f_0 \leq f_{M,\lambda}}\D m_1\D m_2
g_{\CO}(m_1)g_{\CO}(m_2)\;\times\\
&&\,  \nu^{\II}_1(m_1,m_2)
\frac{1-2(f_0/f_{M,\max})^{2/3}}{\left[1-(f_0/f_{M,\max})^{2/3}\right]^2}
\left(\frac{f_0}{f_{100}}\right)^{-2/3} \frac{\D
f_0}{f_0},\label{eq:dnu2rel}
\end{align}
In Eqs.~(\ref{eq:dnu1rel}) and (\ref{eq:dnu2rel}) $\nu^{\I}_1$ and
$\nu^{\II}_1(m_1,m_2)$ are the non-relativistic terms given by
(\ref{eq:nu1I}) and (\ref{eq:nu1II(m1,m2)}).

Fig.~\ref{fig:latticerate} shows the resulting total event rates for
$\D \ln f_0$ intervals. The solid and dashed lines represent the
total event rates of PEs for Models I and II, respectively,
including the relativistic correction for encounters that avoid
collisions. As a comparison, dotted lines display non-relativistic
results. Compared to the non-relativistic results, event rates
decrease for two reasons: first the gravitational focusing decreases
the cross sections of relativistic orbits, for
$2100\Hz=f_{0,\max}[2m_{\NS},(\lambda=2)]\gsim f_0\gsim
f_{0,\max}[2m_{\max},(\lambda=6)]=10\Hz$, and second, the plunging
orbits with $\lambda<2$ are excluded from our estimate. The latter
effect kicks in at $f_{0,\max}(2m_{\max},2)>47\Hz$ where the highest
mass BHs suffer head-on collisions. At
$f_{0,\max}(2m_{\min},2)=570\Hz$ even the smallest BHs are captured,
and only the NS--NS PE event rate contributions remain. The NS--NS
partial event rates can be visualized for lower frequencies by
extrapolating the total event rates shown in
Fig.~\ref{fig:latticerate} between $570\Hz<f_0<2100\Hz$. The NS--NS
event rates are clearly negligible compared to the total rates
including BHs. Note, that our calculations use point masses valid
for BHs only. For $f_0\gsim 1500\Hz$ the minimum separation of
$1.35\Msun$ NSs decreases under $\sim20\km$, for which our
approximation breaks down.

Figure~\ref{fig:latticerate} is useful to visualize the total PE
event rate per GC. However, only a fraction of these events can be
detected, and this fraction depends on both the differential
encounter event rates $\partial^3\nu/\partial m_1\partial
m_2\partial \ln f_0$ and also the observable distance of the
encounter. For the detection rates we shall make use of the
infinitesimal encounter event rate for infinitesimal mass and $f_0$
bins. From eq.~(\ref{eq:dnu2rel}) we get\footnote{We follow to
notation of \cite{mil02} and \cite{w04} for the definition of
partial event rates by not including the mass distribution
$g_{CO}(m)$. The mass distributions enter only when integrating for
the total event rates eq.~(\ref{eq:rate}).}
\begin{equation}\label{eq:dnu/dm1/dm2/dlnf}
\frac{\partial^3\nu}{\partial m_1\partial m_2\partial \ln f_0} =
\nu^{\II}_1(m_1,m_2)
\frac{1-2(f_0/f_{M,\max})^{2/3}}{\left[1-(f_0/f_{M,\max})^{2/3}\right]^2}
\left(\frac{f_0}{f_{100}}\right)^{-2/3}.
\end{equation}
The {\it total event rate} for one GC depends only on
$\partial^3\nu/\partial m_1\partial m_2\partial \ln f_0$, thus in
eq.~(\ref{eq:dnu2rel}) eq.~(\ref{eq:dnu/dm1/dm2/dlnf}) was directly
integrated over $f_0$, and the $m_1$, and $m_2$ distributions.
However for the {\it detection rate}, the observation distance of
the encounter depends on $f_0$, $m_1$, and $m_2$ differently.
Therefore the differential {\it detection rate} has a modified
parameter dependence, implying that the integration can be carried
out only after the observation distance had been included in the
differential rate.

\subsection{Maximum Distance of Detection}
\label{sec:sub:dist}

We now derive the maximum detectable distance of an encounter for
fixed masses $m_1$, $m_2$, and characteristic frequency  $f_0$, for
a given signal-to-noise ratio $S/N$. The luminosity distance can be
expressed with the redshifted parameters, $m_{iz}=(1+z)m_i$ for
$i=\{1,2\}$  and $f_{0z}=f_0/(1+z)$, using the angular averaged
signal-to-noise ratio (\ref{eq:snr}) and signal waveform
(\ref{eq:h(f)2}):
\begin{equation}\label{eq:DL}
\dL(m_{1z},m_{2z},f_{0z}) =
\frac{\sqrt{85}\,\pi^{2/3}}{2^{5/3}}\frac{\G^{5/3}}{\cl^4}
\frac{{\Mchirp_z}^{5/3}}{S/N}\sqrt{\frac{E_{\rm
rel}(\lambda)}{E_{\rm nr}(\lambda)}}W(f,f_{0z}),
\end{equation}
where $E_{\rm rel}(\lambda)/E_{\rm nr}(\lambda)$ is the enhancement
of the GW energy for general relativistic orbits \citep[and see
\S~\ref{sec:PE waveforms} above]{gkl05a}, where
$\lambda=\lambda(M,f_0)=\lambda(M_z,f_{0z})$ is given by
eq.~(\ref{eq:lambda}), and $W(f,f_{0z})$ is a factor depending on
only the frequencies
\begin{equation}
W(f,f_{0z}) = \sqrt{\frac{4}{5}\int_{f_{\min}}^{f_{\max}}
\frac{f_{0z}^{4/3}}{f^2}\frac{F(f/f_{0z})}{S_n(f)}\;\D f}
\end{equation}
$F(x)$ is the dimensionless, normed GW energy spectrum defined in
\S~\ref{sec:PE waveforms}, $f_{\min}$ and $f_{\max}$ are the minimum
and maximum frequencies specific for GW detectors (see
\S~\ref{sec:Detectors}). Henceforth we shall fix $S/N=5$, but other
values can be roughly obtained by scaling the final result on
detection rates with $(S/N)^{-3}$ (assuming that the number of
sources increases with ${\dL}_{\max}^3$).

Note that eq.~(\ref{eq:DL}) formally depends on redshifted
parameters. However, since both the differential encounter event
rate (\ref{eq:dnu/dm1/dm2/dlnf}) and the GC model mass distribution
depend on the comoving parameters, it is useful to revert to the
comoving parameters in eq.~(\ref{eq:DL}) and get $\dL(m_1,m_2,f_0)$.
This can be achieved by writing $m_{iz}=(1+z)m_{i}$ for $i=\{1,2\}$
and $f_{0z}=f_0/(1+z)$ in (\ref{eq:DL}) and making this equal to the
standard formula ${\dL}_{\rm cos}(z)$ connecting the luminosity
distance and redshift in a specific cosmology \cite[e.g][the index
``cos'' refers to the cosmological luminosity distance--redshift
formula in order to distinguish this from the maximum distance
(\ref{eq:DL}) specific for PE encounters]{eis97}. Now both sides
depend on $z$. Numerically solving for $z$ gives $z(m_1,m_2,f_0)$.
Finally substituting the result back in ${\dL}_{\rm cos}(z)$ gives
$\dL(m_1,m_2,f_0)$.

This procedure is however cumbersome in practice. It becomes
numerically very time-consuming when computing the total detection
rates, which includes the evaluation of integrals over the
parameters. Therefore we make the following essential approximations
when solving for the luminosity distance in (\ref{eq:DL}):
\begin{equation}\label{eq:DL2}
\dL(m_1,m_2,f_0) \equiv \left\{
\begin{array}{ll}
  \mbox{using~} z=0 & \mbox{if~}z_1\leq 0.01 \\
  \mbox{using~} z={\rm H}_0\dL/\cl & \mbox{if~}z_1\leq 0.1 \\
  \mbox{no approximations} & \mbox{if~} 0.1<z_1<6 \\
  \mbox{using~} z=6 & \mbox{if~} S/N>5\mbox{~for~}z= 6 \\
\end{array}
\right.
\end{equation}
On the RHS of eq.~(\ref{eq:DL2}), $z_1$ is the first approximation
of the redshift, which is obtained by calculating $\dL$ from the RHS
of eq.~(\ref{eq:DL}) with no redshift, and making this equal to
${\dL}_{\rm cos}(z_1)$, and solving for $z_1$. We neglect cosmology
for $z_1<0.01$ and take a Hubble constant ${\rm H}_0=70\km/{\rm
s/Mpc}$ for $0.01<z_1<0.1$. Next, whenever $z_1>0.1$, we substitute
$z=6$ for the RHS of eq.~(\ref{eq:DL}) and in case this is already
larger than ${\dL}_{\rm cos}(6)$, then we conclude that the source
is observable at $z=6$ for $S/N>5$ and take $\dL = {\dL}_{\rm
cos}(6)$ as the maximum distance of observation. We do not explore
detection rates at larger redshifts, since then the BH mass and
radial distribution might not have relaxed to the final state. If
$S/N< 5$ for $z=6$ then we execute the exact procedure without
approximations for $z_1>0.1$

Changing to the non-redshifted variables in Eq.~(\ref{eq:DL}) is
even ambiguous in some cases. For {\it NGLISA} the signal-to-noise
ratio is occasionally not a decreasing but an increasing function of
the redshift. This happens when the signal is redshifted to the more
sensitive range of frequencies of the detector, and the enhancement
in sensitivity is more substantial than the attenuation from
increasing the distance. In these cases a certain encounter can be
observed within a certain distance, then increasing the distance,
the encounter first becomes invisible (i.e. $S/N<5$), then again
visible (i.e. $S/N\geq 5$) within a second maximum distance. This
phenomenon occurs for {\it NGLISA} for large BH masses and
near-maximum characteristic frequencies.

\begin{figure}
  \centering\mbox{\includegraphics[width=8cm]{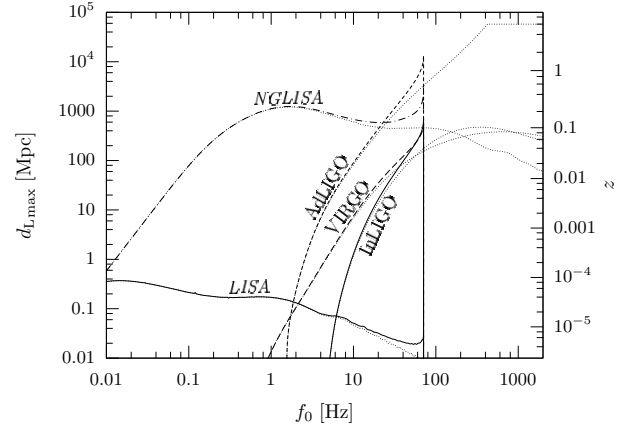}}
  \caption{\label{fig:Dmax} Maximum luminosity distance,
  ${\dL}_{\max}$, of two BHs with $m_1=m_2=40\Msun$ masses undergoing
  a PE, and emitting GWs that are detected on average with $S/N=5$.
  The $x$-axis shows the emitted $f_0$ characteristic frequency
  of the flyby comoving with the host GC. The curves correspond
  to InLIGO, VIRGO, AdLIGO, {\it LISA},
  and {\it NGLISA}, respectively. Thin dotted lines show the result
  for Newtonian waveforms \citep{Turner}, thick lines account for
  general relativistic corrections to the GW amplitudes
  for close encounters. For
  frequencies larger than $f_{M,\max}=71\Hz$ the minimum distance is
  under $\lambda=2$ Schwarzschild radii, for which a head-on collision takes
  place. We did not impose any restrictions on $\lambda$ for the
  non-relativistic curves. All curves account for the redshifting $z$,
  which is shown on the right border. We restrict to $z\leq 6$. For
  different masses, $\dL$ scales with roughly $\Mchirp^{5/3}$ and the
  cutoff frequency scales with $M^{-1}$. Since signals are broadband,
  the detectors have a chance to observe a broad range of $f_0$.}
\end{figure}
Fig.~\ref{fig:Dmax} shows the maximum distance of sources
with $S/N=5$ for $m_1=m_2=40\Msun$ BH masses. The thick curves
account for the relativistic corrections with minimum separations
larger than $\lambda=2$, the thin dotted lines represent the
non-relativistic results with no bound on $\lambda$. The maximum
frequency (\ref{eq:fMmax}) corresponding to marginally colliding
orbits ($\lambda=2$) is $71\Hz$ and scales with $M^{-1}$ for other
masses. For other masses ${\dL}_{\max}$ scales with $\Mchirp^{3/5}$.
All curves account for the cosmological redshifting. The plot shows,
that the non-relativistic approximation is adequate for small
frequencies, but it implies a luminosity distance a factor of 2 -- 3
lower than the relativistic calculation near the maximum $f_0$
frequency. Therefore, the dotted lines are useful to approximately
visualize ${\dL}_{\max}$ for lower $M$, when the cutoff frequency
shifts to higher values.

The enclosed volume and the observable sources are given by
eq.~(\ref{eq:N_gccos}). If neglecting relativistic and cosmological
effects, we get $V \propto D^3\propto {\cal M}^5$.

\subsection{Detection Rates}
\label{sec:sub:rate3}

In the previous sections we calculated the differential event rates
of PEs for single GCs per infinitesimal mass and frequency bins, and
computed the maximum distance of their detection. Here we combine
these results to calculate the total detection rate of PEs.

For fixed $m_1$, $m_2$, and $f_0$, the rate of GW detections of the
corresponding encounters is the observed rate for a single GC times
the number of observable GCs. Since there is a cosmological redshift
between the source GC and the observation, the single-GC rate is
reduced by $1+z$:
\begin{equation}\label{eq:dnu_totdef}
\frac{\partial^2\nu^{\rm total}}{\partial m_1\partial m_2\partial
\ln f_0} = \frac{1}{1+z}\frac{\partial^2\nu^{\rm single}}{\partial
m_1\partial m_2\partial \ln f_0} N^{gc}(m_1,m_2,f_0).
\end{equation}
The first term is the redshifted event rate expressed with the
comoving event rate (\ref{eq:dnu/dm1/dm2/dlnf}) and the second is
simply
\begin{equation}
N^{\gc}(m_1,m_2,f_0) = N^{\gc}[{\dL}_{\max}(f_0,m_1,m_2)],
\end{equation}
where $N^{\gc}(\dL)$ is the number of GCs within a given maximum
luminosity distance. In practice ${\dL}_{\max}$ is given by
eq.~(\ref{eq:DL2}) which we substitute in eq.~(\ref{eq:N_gccos}).

The total detection rate is then simply the integral of the
differential detection rate (\ref{eq:dnu_totdef}) using the CO mass
distribution. After substituting, we get
\begin{eqnarray}
\nu^{\rm tot}(f_0)& =& \int_0^{f_{0\max}} \frac{\D f_0
}{f_0}\int\hspace{-1ex}\int_{f_0 \leq f_{M,\max}}\D m_1\D m_2 \nonumber\\ &&\quad
g_{\CO}(m_1)\,g_{\CO}(m_2)\, \frac{\partial^2\nu^{\rm
total}}{\partial m_1\partial m_2\partial \ln f_0} \label{eq:rate}
\end{eqnarray}
where $g_{\CO}(m)$ is the CO mass distribution for Model II, defined
in \S~\ref{sec:sub:sub:Model2}. The mass integrals are evaluated
over the $(m_1,m_2)$ domain for which the encounter avoids a
collision (i.e. $f_0 \leq f_{M,\max}$, see eq.~[\ref{eq:fMmax}]) and
the $f_0$ integral extends to a maximum possible frequency
independent of masses ($\sim f_{2m_{\NS},\max}$). The result of
eq.~(\ref{eq:rate}) is one number, the expected rate of detection
for the specific detector.

\subsection{Results}
\label{sec:sub:rateresult}

The estimated total number of successful detections from
eq.~(\ref{eq:rate}) is $\nu^{\rm tot}=5.5\times 10^{-5}\yr^{-1}$ for
InLIGO, $7.2\times 10^{-5}\yr^{-1}$ for VIRGO, $6.3\times
10^{-2}\yr^{-1}$ for AdLIGO, $2.9\times10^{-6}\yr^{-1}$ for {\it
LISA} and $1.0\yr^{-1}$ for {\it NGLISA}.

\begin{figure}
  \centerline{\includegraphics[width=8cm]{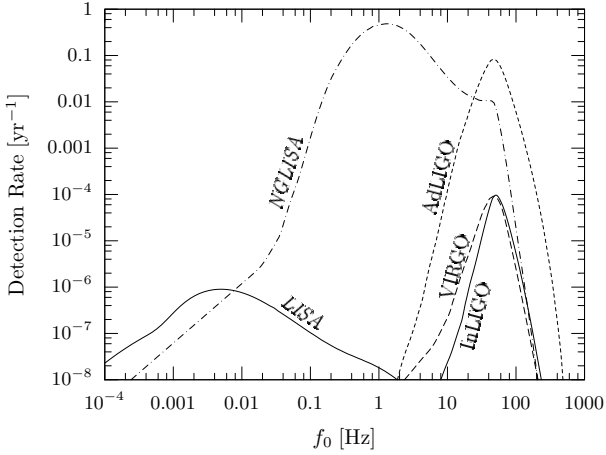}}
  \caption{\label{fig:rate} The expected PE detection rate
  per logarithmic emitted characteristic frequency bin.
  Results are shown for the specific detectors. All curves account for
  general relativistic encounters and cosmology. The units on the
  $y$-axis is simply $\yr^{-1}$, since it is a rate per $\D \ln f_0 = \D f_0/f_0$
  bins for which the units of $f_0$ drops out. For large $f_0$, there
  is an abrupt cutoff in the detection rate as the
  larger mass BHs suffer head-on collisions, leaving only the
  contribution of low-mass BHs in the PE rate.
  }
\end{figure}

\begin{figure}
  \centerline{\includegraphics[width=8cm]{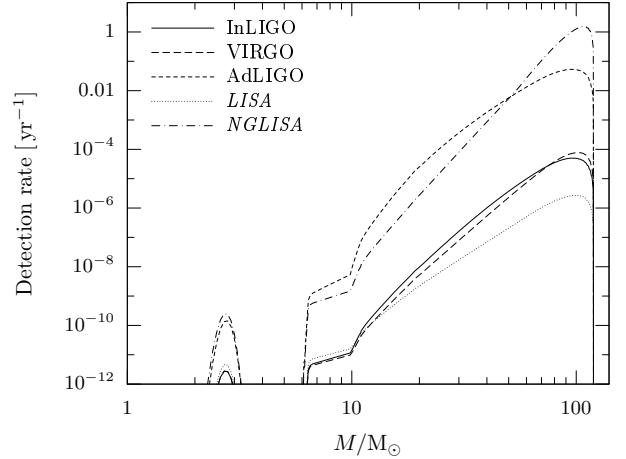}}
  \caption{\label{fig:ratemass}  The expected PE detection rate
  per logarithmic total mass bins, $\D \ln M$, for the various
  detectors. Note that the mass function of our GC model
  is constant for $\D \ln m$ intervals for BHs $120\Msun > 2m >10\Msun$, and
  is Gaussian type for NSs $2m\sim 2.7\Msun$. BH--NS encounters
  dominate for $6.35\Msun<M<10\Msun$. The dominant PE contribution
  is expected from $m_{1,2}=40-60\Msun$ component masses.
  }
\end{figure}

\begin{figure}
  \centerline{\includegraphics[width=8cm]{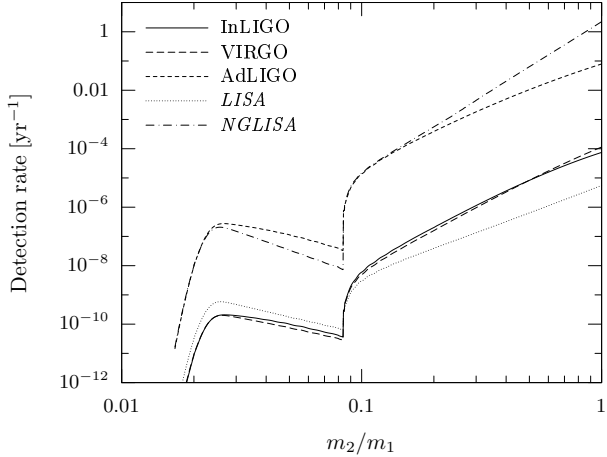}}
  \caption{\label{fig:ratemassratio}  The expected PE detection rate
  as a function of the mass ratio $q=m_2/m_1$ of the interacting
  masses, where $m_1\geq m_2$ is assumed. The partial PE detections
  are plotted per logarithmic mass ratio bins, $\D \ln q$,  for the various
  detectors. Note that the assumed smallest and largest CO masses of our
  GC model implies a cutoff below $1.35\Msun/60\Msun$.
  The detection rate is dominated by equal mass encounters. PEs with
  $q>5\Msun/60\Msun$ are dominated by BH--BH encounters, while
  $1.35\Msun/60\Msun\lsim q<5\Msun/60\Msun$ correspond to BH--NS
  events.
  }
\end{figure}

It is interesting to see the differential event rate per logarithmic
$f_0$ bin independent of masses, which is obtained by carrying out
only the mass integrals in eq.~(\ref{eq:rate}). The result is shown
in Fig.~\ref{fig:rate}. The figure shows that both AdLIGO and {\it
NGLISA} could have some chance to detect PE events, if observing for
one year; AdLIGO mainly sensitive to $f_0$ frequencies between $30$
and $80\Hz$, and {\it NGLISA} sensitive between $0.2\Hz$ and
$10\Hz$. There is a sharp cutoff in the PE detection rate for high
frequencies. In this regime, the encounters among the relatively
higher mass BHs are not parabolic, but result in direct captures,
and only the lower mass BHs contribute to the PE detection rate.
show the differential event rate for logarithmic total mass bins,
$\D \ln M$, and for logarithmic mass ratio bins, $\D \ln q$. Here we
define the mass ratio as $q=m_</m_>$ for which $q\leq 1$. (Recall
the definitions $m_<=\min(m_1,m_2)$ and $m_>=\max(m_1,m_2)$.) The
$M$ dependent partial PE detection rate can be obtained from
eq.~(\ref{eq:rate}) by changing the $m_1, m_2$ integrals to $M$ and
$m_2$ variables, rearranging the order of integrals, and evaluating
the $f_0$ and $m_2$ integrals only. The partial PE rates for fixed
$q$ can be obtained similarly, by changing to $m_>$ and $q$
variables, and evaluating the $f_0$ and $m_>$ integrals only. In
\S~\ref{sec:sub:sub:DerivationGC} we demonstrated that the event
rates of GCs are sensitive to $\langle m^{10/3}\rangle$, and are
inclined towards the high-mass end of the CO distribution, in
particular PEs of NSs have a relatively negligible event rate. The
detectable volume entails an even stronger mass dependence $m^5$.
Therefore, for a mass distribution of $m^{-1}$, we expect a scaling
with $\sim m^{22/3}$ for logarithmic total mass bins, implying that
the highest mass BHs will dominate the PE event rates. However,
increasing the BH masses decreases the maximum $f_0$ frequency of an
encounter avoiding a collision. Figure~\ref{fig:ratemass} verifies
that all of the detectors are indeed much more sensitive to large
total masses, even though our model GC (i.e. Model II) contains a
small relative number of these objects. Note, that the BH mass
distribution $g_{\BH}(m)$ is constant for $\D \ln m$ bins. With
Figure~\ref{fig:ratemass} the partial detection rates of BH--BH and
NS--NS encounters can be visualized. For $M > 2m_{\min}=10\Msun$ the
BH--BH encounters dominate, while $M\approx2m_{\NS}$ correspond
exclusively to NS--NS encounters. PE detections of NS--NS encounters
are practically impossible, they are suppressed by at least 9 orders
of magnitudes. Similarly, Figure~\ref{fig:ratemassratio} shows that
BH--NS encounters are also suppressed by 6 orders of magnitudes!

\begin{figure}
  \centerline{\includegraphics[width=8cm]{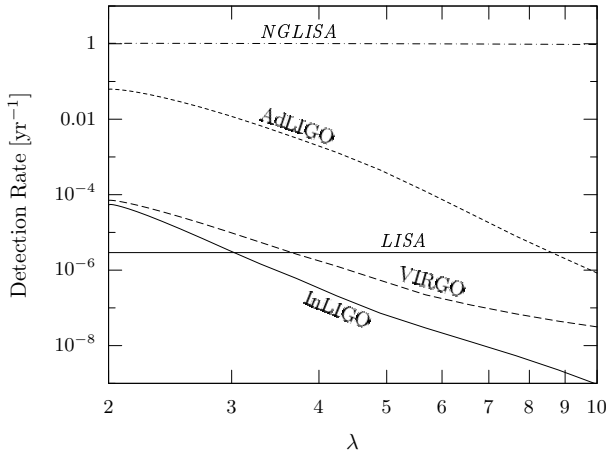}}
  \caption{\label{fig:ratelambda} The total cumulative expected
  detection rate of PEs subject to the constraint that the minimum
  separation exceeds $\lambda$ times the total Schwarzschild radius.
  For $\lambda< 2$ a head-on collision occurs, which we are not
  considering in the present paper (see \S~\ref{sec:sub:comparison}
  for a discussion).
  }
\end{figure}

Figure~\ref{fig:ratelambda} shows the detection rate as a function
of minimum distance, $\lambda_{\min}$ of the encounters. Recall that
for a given total mass $M$, $\lambda$ determines the characteristic
frequency  $f_0$ by eq.~(\ref{eq:fconstraint}), and marginally
plunging orbits correspond to $\lambda_{\min}=2$.
Figure~\ref{fig:ratelambda} was obtained by changing the domain of
integration of $f_0$ to $f_0\lsim f_{M,\lambda}$ in
eq.~(\ref{eq:rate}). The curves show that terrestrial detectors are
more sensitive to close approaches than space detectors. The
$\lambda_{\min}=2$ case corresponds to all of the PE detections. It
is interesting to note, that terrestrial detectors display a
different $\lambda$ dependence: AdLIGO rates show a weaker increase
for marginally colliding orbits $\lambda\sim 2$. This is a
consequence of cosmology: the observation distance is so large
(Fig.~\ref{fig:Dmax}) that the cosmological comoving volume element
is significantly smaller, and the GW frequency is redshifted outside
the sensitive domain of the detector for these events.

\section{Conclusions}
\label{sec:Conclusions}

PEs of solar BHs are possible sources of gravitational radiation.
Our results suggest that current and near future GW detectors are
potentially capable of detecting these signals in the local universe
and up to cosmological distances for the higher masses. We
anticipate $S/N\gsim 5$ matched-filtering detection rates for
quasi-parabolic trajectories avoiding collisions, $\nu^{\rm
tot}=5.5\times 10^{-5}\yr^{-1}$ for InLIGO, $7.2\times
10^{-5}\yr^{-1}$ for VIRGO, $0.063\yr^{-1}$ for AdLIGO,
$2.9\times10^{-6}\yr^{-1}$ for {\it LISA}, and $1.0\yr^{-1}$ for
{\it NGLISA}. For different signal-to-noise ratios, detection rates
scale by approximately $(S/N)^{-3}$. These results correspond to a
BH mass function $g_{\BH}(m)\propto m^{-1}$ with minimum and maximum
masses of $5$ and $60\Msun$. For comparison we ran calculations for
more general distributions $g_{\BH}(m)\propto m^{-p}$, and different
$m_{\min}$ and $m_{\max}$ values. Table~\ref{tab:models} lists the
results for these models. Here, we have fixed $N_{\BH}=500$ for
models with $m_{\min}=5\Msun$. For larger $m_{\min}$, we reduce
$N_{BH}$ by assuming that BHs with masses $5\Msun\leq m<m_{\min}$
have escaped the cluster. Results are very different for various
choices of parameters (see \S~\ref{sec:sub:BHMassDistribution} for a
detailed discussion below).

\begin{deluxetable}{ccclllll} \tablecolumns{8}
 \tablecaption{\label{tab:models}Detection rates for alternative models}
 \tablehead{\colhead{$p$} & \colhead{$m_{\max}$} & \colhead{$m_{\min}$} &
 \colhead{$\nu_{\rm InLIGO}$} & \colhead{$\nu_{\rm VIRGO}$} & \colhead{$\nu_{\rm AdLIGO}$}
& \colhead{$\nu_{\it LISA}$} & \colhead{$\nu_{\it NGLISA}$}\\
\colhead{} & \colhead{$[\Msun]$} & \colhead{$[\Msun]$} &
 \colhead{$[\yr^{-1}]$} & \colhead{$[\yr^{-1}]$} & \colhead{$[\yr^{-1}]$}
& \colhead{$[\yr^{-1}]$} & \colhead{$[\yr^{-1}]$}}
 \startdata
   $0$  & $20$  & $5$    & $1.0\, (-6)$ & $5.2\, (-7)$ & $2.2\, (-3)$ & $8.7 \, (-8)$ & $6.3\, (-4)$\\
   $0$  & $60$  & $5$    & $2.2\, (-4)$ & $3.1\, (-4)$ & $2.4\, (-1)$ & $1.2 \, (-5)$ & $5.2$\\
   $0$  & $60$  & $40$   & $1.3\, (-4)$ & $2.1\, (-4)$ & $1.3\, (-1)$ & $7.2 \, (-6)$ & $4.3$\\
   $0$  & $100$ & $5$    & $1.0\, (-3)$ & $3.6\, (-3)$ & $1.3$        & $1.3 \, (-4)$ & $ $\\
   $0$  & $100$ & $40$   & $9.5\, (-4)$ & $3.4\, (-3)$ & $1.2$        & $1.2 \, (-4)$ & $ $\\
   $0$  & $100$ & $80$   & $2.4\, (-4)$ & $1.2\, (-3)$ & $3.4\, (-1)$ & $4.3 \, (-5)$ & $ $\\
   $1$  & $20$  & $5$    & $4.8\, (-7)$ & $2.4\, (-7)$ & $1.0\, (-3)$ & $4.5\, (-8)$  & $2.7\, (-4)$\\
 $\bf 1$  & $\bf 60$  & $\bf 5$    & $\bf 5.5\, (-5)$ & $\bf 7.2\, (-5)$ & $\bf 6.3\, (-2)$ & $\bf 2.9\, (-6)$  & $\bf 1.0$\\
   $1$  & $60$  & $40$   & $2.5\, (-5)$ & $3.9\, (-5)$ & $2.6\, (-2)$ & $1.3\, (-6)$  & $7.6\, (-1)$\\
   $1$  & $100$ & $5$    & $2.2\, (-4)$ & $6.4\, (-4)$ & $2.8\, (-1)$ & $2.4\, (-5)$  & $ $\\
   $1$  & $100$ & $40$   & $1.8\, (-4)$ & $5.7\, (-4)$ & $2.2\, (-1)$ & $2.0\, (-5)$  & $ $\\
   $1$  & $100$ & $80$   & $2.9\, (-5)$ & $1.4\, (-4)$ & $4.2\, (-2)$ & $5.2\, (-6)$  & $ $\\
   $2$  & $20$  & $5$    & $1.8\, (-7)$ & $9.1\, (-8)$ & $4.1\, (-4)$ & $2.1\, (-8)$  & $9.5\, (-5)$\\
   $2$  & $60$  & $5$    & $6.6\, (-6)$ & $7.2\, (-6)$ & $8.3\, (-3)$ & $3.7\, (-7)$  & $8.4\, (-2)$\\
   $2$  & $60$  & $40$   & $1.8\, (-6)$ & $2.8\, (-6)$ & $1.9\, (-3)$ & $1.0\, (-7)$  & $5.0\, (-2)$\\
   $2$  & $100$ & $5$    & $1.7\, (-5)$ & $3.6\, (-5)$ & $2.1\, (-2)$ & $1.5\, (-6)$  & $ $\\
   $2$  & $100$ & $40$   & $9.7\, (-6)$ & $2.7\, (-5)$ & $1.1\, (-2)$ & $9.6\, (-7)$  & $ $\\
   $2$  & $100$ & $80$   & $8.9\, (-7)$ & $4.3\, (-6)$ & $1.3\, (-3)$ & $1.6\, (-7)$  & $ $
 \enddata
\tablecomments{The number of BHs per GC is normalized to
$N_{\BH}=500$ for $m_{\min}=5\Msun$ for all choices of $m_{\max}$.
For larger $m_{\min}$, all BHs with $5\Msun\leq m<m_{\min}$ are
assumed to have escaped from the cluster. Detection rates are given
in normal form, where the exponent is shown in parenthesis. Some
fields left blank correspond to cases where the $S/N$ is not a
monotonically decreasing function of distance (see
\S~\ref{sec:sub:dist}).}
\end{deluxetable}

We constructed two different GC models. We conclude that a uniform
mass and density distribution (Model I) is inadequate since the
contribution of the GC core consisting of the more massive BHs are
significantly underestimated. After accounting for mass distribution
and mass segregation, as well as the relative velocity distribution
of the sources (Model II) we obtained event rates two orders of
magnitudes higher than Model I per GC. Moreover, more massive BHs in
GCs are visible to significantly larger distances, and supply the
most prominent sources of PEs for detection
(Fig.~\ref{fig:ratemass}).

In \S~\ref{sec:Discussion}, we include a critical review of our
assumptions and their influence on the results. To point out just
one thing, note that compared to our previous estimates above, the
PE detection rates might have been underestimated by four orders of
magnitudes for core-collapsed GCs (depending on the final core
radius and population, see \S~\ref{sec:sub:CoreCollapse} below)!

\section{Discussion}\label{sec:Discussion}
\subsection{Comparison with Other Orbits}\label{sec:sub:comparison}

How do PE event rates compare to the event rates of binary
inspirals, mergers, and head-on collisions? What are the main
factors for the difference? We briefly discuss these questions in
this section.

\subsubsection{Basic Features of Parabolic Encounters}
Let us quickly summarize the main properties of PE sources.
\begin{itemize}
  \item The event rates for $\D \ln f_0$ intervals scale with $f_0^{-2/3}$
for trajectories avoiding collisions. Collisions decrease PE event
rates quickly for large frequencies, $f\gsim 50\Hz$, rates drop four
orders of magnitudes between 50 and $500\Hz$.
  \item The $f_0$-scaling of the amplitude of the signal is
$f_0^{-1/3}$ for GW frequencies $f\sim f_0$. The integrated RSS
signal amplitude scales with $f_0^{1/6}$.
  \item The signal energy-spectrum is broadband, has a maximum at $f\sim
f_0$, and a relatively shallow cutoff for larger frequencies. The
half-maximum of the signal is $\Delta f\sim 1.5f_0$, and the
spectral energy density drops to $1\%$ at $f\sim 5f_0$
\citep{Turner}.
  \item In terms of detections for $\D \ln f_0$ intervals, the maximum
distance of PE detections for a band-pass detector is roughly
independent of frequencies between $f_{\min}\lsim f_0\leq
f_{0\max}$, where $f_{\min}$ is the minimum detectable frequency of
the detector and $f_{0\max}$ is the maximum characteristic frequency
of a PE avoiding collisions.
  \item The detection rates of equal-mass PEs scale with
$\langle m^{22/3}\rangle$ for $\D \ln m$ intervals, the higher mass
BHs dominate PE detections.
  \item Space detectors will possibly detect more PE
events in the local universe, but terrestrial detectors see further.
  \item Typical event rates are
$1.6\times 10^{-12}\yr^{-1}{\rm GC}^{-1}$ or equivalently ${\cal R}
= 1.4\times 10^{-11}h^3\yr^{-1}\Mpc^{-3}(\Delta \ln f_0)^{-1}$ for
$f_0=50\Hz$.
  \item Typical maximum distance of detection for PEs {\it with
appropriate} $f_0$ is $\sim 300\Mpc$ for InLIGO and VIRGO, $z\sim 1$
for AdLIGO, $\sim 0.4\Mpc$ for {\it LISA}, and $z\sim 0.2$ for {\it
NGLISA}.
  \item Typical overall rate of PE detections per year is
$\sim 10^{-4}$ for InLIGO and VIRGO, $\sim 0.1$ for AdLIGO, $\sim
10^{-6}$ for {\it LISA}, and $\sim 1$ for {\it NGLISA}.
\end{itemize}

Table~\ref{tab:comparison} contrasts the event rates of PEs and
other possibly detectable sources in the literature. The event
rates, $\cal R$, listed in the table are normalized to an average
space density of galaxies $n_{\gal}=0.029\Mpc^{-3}$ and
$n_{\gc}=2.9\Mpc^{-3}$ for GCs (see \S~\ref{sec:sub:GCabundance}).
For PEs the $\cal R$ values shown correspond to the sensitive range
of characteristic frequencies: $10\Hz\leq f_0\leq 200\Hz$ for
InLIGO, VIRGO, AdLIGO, $10^{-4}\Hz\leq f_0\leq 1\Hz$ for {\it LISA},
and $10^{-1}\Hz\leq f_0\leq 60\Hz$ for {\it NGLISA} (see
Fig.~\ref{fig:rate}). The maximum distance of detection, $\dL$,
shown are typical values for detectable encounters. The elements
marked with ``-'' are not given by the corresponding references.
Values are left blank that are the same as in the previous row. Note
that the numbers are very uncertain depending on model assumptions.
PE event rates correspond to our standard GC model, most significant
uncertainties are the numbers of higher mass BHs in the cluster.
Note, that stellar BH--BH inspiral event rate estimates in GCs vary
3 orders of magnitudes! 
\begin{deluxetable}{llllll} \tablecolumns{6}
 \tablecaption{\label{tab:comparison}Comparison of event and detection rates with other sources}
 \tablehead{
 \colhead{Event} & \colhead{Loc} & \colhead{$\cal R$} &  \colhead{Detector} & \colhead{$\dL$} & \colhead{$\nu$}
 \\
 \colhead{} & \colhead{} & \colhead{}
 & \colhead{} & \colhead{$[\Gpc]$} & \colhead{$[\yr^{-1}]$}}
 \startdata
    BH/BH PE\tablenotemark{1}        & GC & $7.1\,(-11)$ & InLIGO  & $3.0\,(-1)$   & $5.5\,(-5)$ \\
              & GC &      $7.1\,(-11)$         & VIRGO   & $3.0\,(-1)$       & $7.2\,(-5)$ \\
              & GC &    $7.1\,(-11)$           & AdLIGO     & $5.6$       & $6.3\,(-2)$ \\
              & GC & $2.4\,(-7)$ & {\it LISA}  & $5.1\,(-4)$ & $2.9\,(-6)$ \\
              & GC &  $2.4\,(-9)$& {\it NGLISA}& $1.9$      & $1.0$ \\
    NS/NS insp\tablenotemark{2}     & Field     & $1.9\,(-7)$ & AdLIGO &  $2.0\,(-1)$ & $3.0$ \\
    \quad(\tablenotemark{3}LPP)     & Field     & $1.1\,(-6)$ & AdLIGO &  $2.0\,(-1)$ & $3.4\,(+1)$ \\
    \quad(\tablenotemark{4}Kalogera)& Field     & $2.0\,(-6)$ & InLIGO &  $2.0\,(-2)$ & $3.5\,(-2)$ \\
                                    &           & $2.0\,(-6)$ & AdLIGO &  $3.5\,(-1)$ & $1.9\,(+2)$ \\
    \quad(\tablenotemark{5}NGF) & Field & $2.9\,(-4)$ & InLIGO  & $2.0\,(-2)$ & $3.4\,(-1)$ \\
                                &       & $2.9\,(-4)$ & AdLIGO  & $3.0\,(-1)$ & $1.1\,(+4)$ \\
    BH/NS insp\tablenotemark{5} & Field & $2.9\,(-4)$ & InLIGO  & $4.3\,(-2)$ & $3.3\,(+1)$ \\
                                &       & $2.9\,(-4)$ & AdLIGO  & $6.5\,(-1)$ & $1.2\,(+5)$ \\
    BH/BH insp\tablenotemark{6}
     & Field     & $2.0\,(-9)$ & AdLIGO &  $1.1$  & $1.0$ \\
    \quad(\tablenotemark{7}PZM)
     & Nucleus    & $1.8\,(-8)$ & AdLIGO &  $1.1$  & $1.0\,(+2)$\\
     & Zero-age GC& $3.2\,(-7)$ & AdLIGO &  $1.1$  & $1.8\,(+3)$\\
     & Evolved GC & $5.4\,(-8)$ & AdLIGO &  $1.1$  & $3.0\,(+2)$\\
     \quad(\tablenotemark{8}Miller) & GC & $3.5\,(-9)$ & AdLIGO &  $1.6$ & $1.0\,(+1)$\\
     \quad(\tablenotemark{9}O'Leary)& Zero-age GC& $8.7\,(-7)$ & AdLIGO & $-$  & $2.7\,(+3)$\\
                                    & Evolved GC & $8.7\,(-10)$ & AdLIGO & $-$  & $2.7$\\
    IMBH/BH\tablenotemark{8}     & GC & $3.5\,(-9)$ & AdLIGO     &  $1.6$ & $4.0\,(+1)$\\
                                 &  & $3.5\,(-9)$   & {\it LISA} &  $2.0\,(-1)$ & $7.0\,(-3)$ \\
     \quad(\tablenotemark{10}Will)& GC & $8.5\,(-9)$ & {\it LISA} &  $4.0\,(-2)$ & $1.0\,(-6)$ \\
     \quad(\tablenotemark{11}GMH) & GC & $5.1\,(-5)$ & {\it LISA} &  $4.0\,(-2)$ & $6.0\,(-3)$ \\
     \quad(\tablenotemark{12}HPZ) & Field & $1.4\,(-7)$ & {\it LISA} &  $3.5\,(-1)$ & $8.6$\\
    SMBH/WD\tablenotemark{13}   & Nucleus & $1.7\,(-8)$  & {\it LISA} &  $1.6$ & $9.4\,(+1)$ \\
     \quad(\tablenotemark{14}HA)& Nucleus & $9.0\,(-8)$  & {\it LISA} &  $1.0$ & $1.3\,(+2)$ \\
    SMBH/BH\tablenotemark{13}   & Nucleus & $3.2\,(-9)$  & {\it LISA} &  $6.6$ & $1.1\,(+3)$ \\
    SMBH/IMBH\tablenotemark{13} & Nucleus & $2.9\,(-12)$ & {\it LISA} &  $6.6$ & $1.0$ \\
     \quad(\tablenotemark{15}PZ)& Nucleus & $8.3\,(-10)$ & {\it LISA} &  $4.4$ & $1.0\,(+2)$
 \enddata
\tablecomments{Event rates $\cal R$ are is in units $h^3{\rm
yr}^{-1}{\rm Mpc}^{-3}$.\\
$^{1}$ Parabolic encounters for our standard GC model. The maximum
distance of detection corresponds to $m_1=m_2=50\Msun$, $f_0=50\Hz$
for InLIGO, VIRGO, AdLIGO and
{\it NGLISA}, and $f_0=10\mHz$ for {\it LISA}.\\
$^{2}$ NS/NS inspirals, \cite{ps96} theoretical models of binary
evolution calibrated to the observed supernova rate.\\
$^{3}$ \cite{lpp97}\\
$^{4}$ \cite{kal04}, based on 3 highly relativistic radio pulsars.\\
$^{5}$ \cite{ngf05}, based on recent observations of 4 short/hard
gamma-ray bursts, identified with NS/NS or BH/NS inspirals.
\\
$^{6}$ BH/BH inspirals, \cite{py98}.\\
$^{7}$ \cite{pzm00}. Distance corresponds to the inspiral of masses
$m_1=m_2=10\Msun$ and detection rates correspond to a $100\%$
detection efficiency
within this distance.\\
$^{8}$ \cite{mil02}\\
$^{9}$ \cite{O'Leary05}. They compute the number of mergers as a
function of time, as BHs are being slowly depleted from the cluster.
The detection rates correspond to cases, where all GCs are zero-age
(1Myr) or evolved (10Gyr), with a number density $1\Mpc^{-3}$.\\
$^{10}$ Stellar BH inspirals into intermediate mass BHs, adopted
from \cite{w04}. Distance corresponds to the inspiral of masses
$m_1=100\Msun$ and $m_2=10\Msun$, $\cal R$ corresponds to the
inspirals with a maximum time $40$--$400\yr$ before merger which can
be detected with
$S/N=10$ after a $1\yr$ integration with {\it LISA}.\\
$^{11}$ \cite{gmh05} assuming more optimistic probability of IMBHs
in
GCs and IMBH maximum mass than \cite{w04}\\
$^{12}$ \cite{hp05}, assuming observed ultraluminous X-ray sources
are associated to tidal captures of $10\Msun$ BHs by $10^3\Msun$ IMBHs.\\
$^{13}$ Inspirals into supermassive BHs, $M=10^6\Msun$. \cite{gair04} assuming their optimistic set of results. \\
$^{14}$ \cite{ha05}\\
$^{15}$ \cite{pz06} N-body simulations of the formation and inward
migration of IMBHs in the Galactic center.}
\end{deluxetable}
\subsubsection{Parabolic Encounters vs. Inspirals}

The event rates of PEs depend on the characteristic frequency.
Without any specifications PEs are more regular than e.g. BH--BH
inspirals. However, {\it within the sensitive range} of GW
terrestrial detectors PEs are rather rare by a factor of
$\sim10^{-4}$. Event rates are higher for space detector
frequencies, however space detectors have a smaller distance of
maximum observation.

Table~\ref{tab:comparison} suggests that the detectable distance of
PEs is comparable to inspirals. However, this is somewhat deceptive
since PE results correspond to larger BH masses, $50\Msun$ rather
than $10\Msun$ which is regular for BH--BH inspirals in the
literature. For $10\Msun$ component masses, the maximum distance of
observation is less for PEs than for inspirals
(Fig.~\ref{fig:Dmax}), the PE event rates are suppressed by a factor
of $\sim 10^{3}$ (Fig.~\ref{fig:ratemass}). However for larger
masses the comparison changes with the following factors. First,
although the GW signal amplitude is proportional to $\Mchirp^{5/3}$
for inspirals, increasing the masses reduces the signal frequency
(which in turn reduces the detector sensitivity) and also reduces
the observation time (which also decreases the effective signal
amplitude). Another important difference is in the binary separation
$\lambda$ which determines the signal frequency. It is restricted to
$\lambda>3$ for inspirals, the innermost stable circular orbit, a
more stringent constraint than the condition $\lambda>2$ for PEs
(see \S~\ref{sec:sub:relativistic}). Finally opposed to the PE
signal waveforms, the inspiral signals are narrow band, implying
that the high mass, i.e. low frequency, inspiral waveforms are much
harder to detect as much smaller signal power accumulates at the
more sensitive range of frequencies. When combining all of these
effects we expect that low mass BH inspirals are detectable further
with terrestrial detectors, while for large masses where the
observation is limited to at most a few orbits, marginally
collisional PEs are better detected. Therefore, detectors can
observe the higher mass encounters for PEs. This is exactly
analogous to the comparison of the inspiral and plunge phases of
binary coalescence, for which the detection of plunge dominates for
large masses \citep{fh98}.

Among the GW detection candidate sources in GCs, PEs are very
infrequent compared to stellar BH--BH inspiral rate estimates of
\cite{pzm00} or \cite{mil02} within GCs, but are comparable to the
recent results of \cite{O'Leary05}. Observations of radio pulsars
and gamma ray bursts suggest several orders of magnitude larger
numbers for NS--NS or BH--NS inspiral detections
\citep{kal04,ngf05}.

\subsubsection{Parabolic Encounters vs. Head-on Collisions}

As a second example, let us consider the event rates of head-on
collisions for unbound encounters. Head-on collisions are related to
PEs, by extending the parameter $\lambda$ to values less than 2, the
unstable circular orbit. Thus it would be relatively straightforward
to extend the analysis to these events, by examining the event rates
for small initial impact parameters, and computing the detectability
as a function of this parameter. However, the exact shapes of
gravity waveforms are presently not available for collisions (see
\citealt{bak06} for current progress), therefore maximum likelihood
detections are not possible, and the detection of these bursts
requires much higher signal-to-noise levels. We shall argue that
direct head-on collision detections are potentially less frequent
than PEs.

The rate of head-on collisions between BHs is well known
\cite[e.g][]{hd76,ckp94,sr97} however the detection rates of the
resultant GW signals is subject to the uncertainty of the GW signals
\citep{fh98}. Direct collisions produce potentially less intensive
GW signals than close PEs even if neglecting the relativistic
amplitude enhancement for PEs. To see this, let us compare GW signal
strengths that we adopt for PEs \citep[see \S~\ref{sec:PE
waveforms}, and ][]{Turner,mar04,gkl05a}, with general relativistic
calculations for head-on collisions. To our best knowledge, off-axis
collisions of BHs have not been calculated as a function of impact
parameter. For radial head-on collision of Schwarzschild BHs $\Delta
E \simeq 0.01\,(\mu^2/M)\,\cl^2$ \citep{d71,ann93,mor99,sper05}.
\cite{sn82} derived GW energies for the radial infall of a test
particle into a Kerr BH, and \cite{mst96} accounted for the spin of
the infalling particle in addition. Results are in the range of
$\Delta E = 0.03$ -- $0.01\,(\mu^2/M)\, \cl^2$ according to the
magnitude and alignment of spins and the relative direction of the
approach. For high-velocity head-on collisions, there are
significantly larger results: $\Delta E = 0.328\,\mu \cl^2$ for
non-rotating BHs \citep{dp92}, and up to $\Delta E = 0.70\,\mu
\cl^2$ for extreme Kerr-BHs \citep{cl03}. However, in GCs the
initial velocities are typically non-relativistic, therefore we do
not expect a significant relative contribution of relativistic
head-on collisions. Once the BHs are so near that a common
surrounding horizon envelope forms the space-time relaxes to a
Kerr-BH. The energy output of this process is between $\Delta E=6
\times 10^{-6}M \, \cl^2$ \citep{pp94} for axisymmetric encounters
and $\lsim 0.01 M \, \cl^2$ for quasi-circular initial conditions
\citep{khanna99}. In comparison, the energy output in GWs for
non-relativistic PEs is $\Delta E_{\rm PE} = 0.01\,
(\lambda/4.1)^{-7/2}\,(\mu^2/M)\, \cl^2$ \citep{Turner}. Using the
low-velocity case, the GW amplitudes of BH collisions are
overestimated by the Newtonian results by a factor between
$(\lambda/3)^{-7/2}$ and $(\lambda/4)^{-7/2}$, depending on spins.
Therefore, the extapolation of the Newtonian treatment to the regime
where the minimum separation is $\lambda\ll 3$ leads to significant
overestimates of the true head-on collision GW energies. In
conclusion, the extrapolation of event rates as a function of
$\lambda$ (for $\lambda<2$ in Fig.~\ref{fig:ratelambda}) or as a
function of the logarithmic characteristic frequency (for
$f_{0}>f_{\lambda=2}$ in Fig.~\ref{fig:Dmax}, dotted lines) is
possibly overly optimistic and therefore inadequate for the
estimation of the detection rates of head-on collisions with the
particular GW detectors.

\subsection{Approximations in the Analysis}\label{sec:sub:approximations}

Our event rate estimates rely on several approximations. The most
important caveat in our analysis is possibly neglecting GW recoil
capture in bound eccentric orbits. The GW radiation reaction is
substantial for strong gravitational fields, for low $\lambda$. For
initially nearly parabolic orbits, the periastron distance and the
eccentricity is decreased\footnote{unless $\lambda\gsim 2.05$, in
which case the eccentricity is increased by GW recoil}
\citep{ckp94}. The first consequence is a minor decrease in the PE
event rate, because of the increase of the cross-section of direct
capture. On the other hand, GW recoil produces bound orbits from
initially unbound trajectories \citep{lee93}. The periastron
distance is then further decreased during each subsequent close
approach inducing successively stronger GW radiation. Therefore PE
events are potential precursors of multiple subsequent more intense
highly eccentric bound encounters, analogous to the captures of
stellar compact objects by supermassive black holes \citep{ha05}.
The GW detection rate of the resultant orbits is likely to be
significantly higher than PE detections. As a result, we anticipate
several successful detections for AdLIGO per year for a wide range
of BH mass-distribution models (see Tab.~\ref{tab:models} for PEs
without GW recoil capture). We leave a detailed quantitative study
for a future paper.

There is a second independent reason suggesting that higher
detection rates will be more likely. Throughout the paper, we
estimated matched filtering detection signal-to-noise amplitudes
with the angular averaged formula which is valid for an analysis
using only a single GW detector. However, a coincident analysis
incorporating several detectors allows much more optimistic
detection limits \citep[see][and \S~\ref{sec:PE waveforms}
above]{jkkt96}. If lowering the angular averaged minimum detection
limit to $S/N=3/\sqrt{5}$ (equivalent to an optimal-orientation
single-detector observation at $S/N=3$) yields $\nu^{\rm
tot}=1.7\times 10^{-3}\yr^{-1}$ for InLIGO, $2.7\times
10^{-3}\yr^{-1}$ for VIRGO, $0.46\yr^{-1}$ for AdLIGO, and
$9.5\times10^{-6}\yr^{-1}$ for {\it LISA} for our standard GC Model
II with $N_{\BH}=500$, $p=1$, $m_{\min}=5\Msun$, and
$m_{\max}=60\Msun$. For models with larger $m_{\max}$ detection
rates are even higher (see Tab.~\ref{tab:models}), implying several
successful detections per year for advanced terrestrial detectors.

We have also neglected the bound binary interactions in the
scattering dynamics and restricted only to single-single encounters.
Depending on the angle of injection this could increase or decrease
event rates. However our analysis of single--single interaction
shows (\S~\ref{sec:sub:sub:DerivationGC}), that the typical
numerical values for PE cross-sections are extremely small, so that
the injection has to have an initial velocity very accurately
pointed towards the target CO in order to produce a detectable
signal. For typical encounters with $m_1=m_2=50\Msun$ and
$f_0=50\Hz$, the minimum separation is $b_0\sim 10^{-6}\AU$, and the
impact parameter is $b_{\infty}\sim 10^{-2}\AU$. As a result, for
single--binary interactions we speculate that the separation of
scales is possible to distinguish three independent phases of the
interaction: (i) the faraway zone $r\gg a_{\rm bin}$, (ii) the
intermediate zone $r\sim a_{\rm bin}$, and (iii) the PE zone $r\sim
b_0\ll a_{\rm bin}$. In (iii) the binary companion can be discarded.
Moreover, note that the velocity during (ii) is still negligible
compared to (iii). Therefore in practice, the beginning of phase
(iii) is exactly analogous to the initial conditions of a
single--single encounter. The only difference is the distribution of
velocities is not isotropic, but after phase (i) it is beamed toward
the center of mass, and phase (ii) adds a random deflection due to
the companion. Plugging in the numbers for binary separations of
$a_{\rm bin}\gg 10^{-2}\AU$ we conclude, binary focusing is not
likely to significantly modify our PE rate estimates. Numerical
simulations would be needed to determine the exact modifications in
the estimates.

Binary interactions also alter the total number and mass
distribution of BHs in the cluster. However in our calculations the
total number and mass function of BHs are input parameters, which
can be chosen consistently with the most sophisticated simulations.

Throughout our analysis we assumed simplified GC models. While our
most sophisticated model accounts for the mass distribution, mass
segregation, and relative velocities (see \S~\ref{sec:sub:GCmodels})
it does not consider the nonuniform radial distribution of density
of regular stars in the cluster core, nor does it consider
variations around the characteristic GC model parameters (e.g.
virial radius, total mass, etc). However, the final results are
simple powers of the characteristic parameters ($\nu\propto q^2
N_{\rm tot}^{1.5} R^{-2.5}$). Our treatment allows upper and lower
bounds to be made on the exact GC model detection rates. These
bounds are still much tighter than other sources of uncertainties,
which justifies the simplifying model assumptions in this analysis.

Another major approximation was to adopt the angular averaged signal
waveforms in the Newtonian approximation \citep{Turner}, and
corrected for the relativistic enhancement of the amplitude,
substantial for close-encounters. We adopted the relativistic
correction for the quadrupole radiation of a test particle geodesics
\citep{mar04,gkl05a} and extrapolated results for other masses.
These estimates do not account for GW recoil. However,
Fig.~\ref{fig:ratelambda} shows that the contribution of extreme
zoom-whirl orbits $\lambda\approx 2$ does not ruin our estimates,
since the detection rate does not increase substantially for
marginally plunging orbits. GW recoil reduces PE signal-power by
driving the interacting masses to collisions, thereby terminating
extreme zoom-whirl orbits much sooner than the no-recoil encounter
time \citep{gkl05b}. We conclude that neglecting GW recoil did not
lead to a large overestimate, implying that our results are
acceptable approximations in this respect.

An exact treatment would have to utilize the more exact
post-newtonian waveforms of the general problem using arbitrary
masses and spins, and should take into account the forward peaking
of GWs for high velocities, Doppler shift of GW frequencies,
spin-orbit, and spin-spin interactions, etc. Although it is clear
that a real data analysis matched filtering would have to be carried
out with exact signal templates, the leading-order (i.e. Newtonian)
term dominates the angular averaged signal power, which is therefore
an adequate first estimate for the detection rates.

\subsection{Uncertainties in the
Result}\label{sec:sub:uncertainties}

\subsubsection{Model Parameters}

There are several uncertainties in our estimate. Among the most
important uncertainties are the values of the GC model parameters,
like the number of BHs in the cluster $N_{\BH}$. \cite{pzm00}
derives $q_{\BH}=N_{\BH}/N_{\rm tot}=6\times 10^{-4}$ by using
\cite{Scalo86} initial mass function (IMF) and assumed that every
object more massive than $20\Msun$ up to $100\Msun$ had evolved to a
BH. When using a Salpeter IMF, the result is $q_{\BH}=10^{-3}$
\citep{mil02}, and \cite{kw03} IMF gives $q_{\BH}=1.5\times 10^{-3}$
\citep{O'Leary05}. We adopt the most conservative result of
\cite{pzm00}. However there is a chance that a non-negligible
fraction of the stars have been ejected from the cluster or have
undergone subsequent mergers. Both processes increase the estimate
on the final BH fraction \citep{mil02}. On the other hand dynamical
binary interactions, binary recoil kicks, or GW recoil of BH mergers
can eject BHs, thereby reducing their overall numbers and possibly
also modify the mass-distribution. In fact, a significant portion of
the stellar-mass BH population might be ejected, especially in small
clusters \citep{sh93,pzm00,O'Leary05}. \cite{belcz05} find that for
an initial binary fraction of $50\%$, the retained fraction of BHs
varies between 0.4 and 0.7. In our fiducial calculations we adopted
$q_{\BH}=5\times 10^{-4}$ and $N^{\rm tot}=10^6$. To see the effects
of BH ejection, Tab.~\ref{tab:models} shows results for other
models. For the general case, we provide analytical scalings which
can be readily used in case these parameters are better determined
in the future. For example since $\nu^{\rm tot}\propto N_{\BH}^2$,
for $N_{\BH}=5000$ ($50$) detection rates increase (decrease) by a
factor of 100.

\subsubsection{Black Hole Mass
Distribution}\label{sec:sub:BHMassDistribution}

An even more significant source of uncertainty is the mass
distribution of BHs in the clusters. We have calculated detection
rates for several distributions (Tab.~\ref{tab:models}). Increasing
the $p$ exponent of the distribution, $g_{\BH}\propto m^{-p}$,
decreases the detection rate by a factor of $\sim 5$ for a unit
change in $p$. Changing the maximum mass of the distribution varies
the results even more significantly. Compared to the detection rate
corresponding to $m_{\max}=60\Msun$, for $m_{\max}=100\Msun$
($20\Msun$) we get a $\sim 1$ -- $3$ order of magnitude increase
(decrease) depending on the detectors. We have also tried changing
the minimum mass $m_{\min}$, by assuming that the BHs with masses
$m<m_{\min}$ have escaped from the cluster. Compared to
$m_{\min}=5\Msun$, a value of $m_{\min}=40\Msun$ reduces detection
rates by a factor of $\sim 2$. In the appendix, we provide
ready-to-use formulas for calculating detection rates for other
parameter values.

From the theoretical point of view, simulations of the initial
stellar BH mass function \citep{fk01} result in a maximum mass limit
of $\sim 20\Msun$, but the particular form of the mass function is
very different for various assumptions (fraction of explosion energy
used to unbind the star, stellar winds, mass transfer after helium
ignition, etc). Recent simulations of rapid star evolution assuming
a lower metallicity for the progenitor stars (weaker stellar winds)
appropriate for GCs and including a large fraction of binaries,
collisions, and accretion leading to the mass buildup of BHs imply a
stellar-mass BH distribution with maximum BH masses around
$M_{\max}=60$ -- $100\Msun$ \citep{belcz05}. Simulations of the
subsequent long-term dynamical evolution has been shown to be
sensitive to BH binary and triple interactions \citep{O'Leary05}.
Binary-single body interactions, BH-star collisions, and GW recoil
kick can possibly significantly reduce the low mass BH population
but enhance the mass of the most massive BHs in the cluster. From
the observational point of view, there is yet lacking evidence for
stellar mass BHs with $m>20\Msun$, but this might be accounted for
the low number statistics \citep[a total of 20 X-ray stellar-mass BH
candidates have been identified to date,][]{cas05}.

\subsubsection{Core Collapse}\label{sec:sub:CoreCollapse}

Finally, a considerable uncertainty in the PE detection rates
results from the actual scaling of the mass segregation
relationships. Even in our complicated model we have assumed a
simple mass segregation, based on thermal equipartition among the
specific CO components. This assumption is in fact valid only among
the decoupled high mass components within the core. \cite{spitz69}
has shown that in simple two-component systems consisting of masses
$m_1$ and $m_2$, with $m_1\ll m_2$, global equipartition cannot be
attained if the low-mass component determines the potential
everywhere in the cluster. In this case, the high mass components
become dynamically decoupled from the rest of the cluster, and the
cluster core collapses to a much smaller radius, $R_{\rm core}$.
This picture has been confirmed by numerical simulations for more
general mass functions \cite[and references therein]{wjr00}.
\cite{gfr04} showed that approximate local thermal equipartition is
attained within the core, and velocities follow $v_m = (K
m/m_{\BH})^{-1/2}v_{\rm core}$, where $v_{\rm core}$ is the velocity
dispersion, $m_{\BH}$ is the mass of components in the core, $K$
describes the departure from equipartition, it is a number of order
1. The total time of the collapse and the final magnitude of core
velocities or core radius, depends sensitively on the initial
fraction of binaries. For a single mass cluster \cite{hth06} found
that $0.01\lsim R_{\rm core}/R_{\gc}\lsim 0.1$, larger values valid
for a large fraction of binaries (here $R_{\gc}$ is the half-mass
radius). In contrast our simple mass segregation led to
$R_{50\Msun}=0.14 R_{gc}$, which is a factor 1.4--14 higher. Note,
that the virial theorem implies $v_m\propto R_m$ for a homogeneous
mass distribution. Detection rates scale with
$R_{m}^{-3}v_{m}^{-1}$, and the contribution of $m\sim 50\Msun$
dominated the final results (see Figs~\ref{fig:ratemass} and
\ref{fig:ratemassratio}). Therefore post-core collapse mass
segregation implies detection rates increased by
$(1.4)^4$--$(14)^4$. Thus, in the most optimistic case, we get a
substantial increase in the detection rates, i.e. $2.1\yr^{-1}$ for
InLIGO, $2.8\yr^{-1}$ for VIRGO, $6.6\days^{-1}$ for AdLIGO,
$0.1\yr^{-1}$ for {\it LISA}, and $4.4\hr^{-1}$ for {\it NGLISA}!

On the other hand, if core collapse leads to runaway collisions and
the buildup of a single intermediate mass black hole, while stellar
mass BHs are ejected from the cluster \citep{fgr06}, PE detection
rates might be considerably suppressed after collapse (i.e.
$\nu\propto N^2_{\BH}$). More information on the typical properties
and long-term evolution of core collapsed clusters is needed to make
PE detection rates less uncertain.

\subsection{Implications}\label{sec:sub:implications}

Opening the gravitational-wave window to observe parabolic
encounters of black holes in globular clusters offers a new
possibility to constrain BH mass functions and GC models. Since PEs
are very sensitive to the number of higher mass stellar BHs
(Figs.~\ref{fig:ratemass} and \ref{fig:ratemassratio}), our results
indicate that a regular detection of PE events would provide
accurate limits on the stellar BH mass distribution in GCs. Our
analysis shows that this might be possible with AdLIGO if average
GCs carry at least 500 BHs.

\subsubsection{Galactic Nuclei}

The analysis can be extended for other spherically symmetric systems
using the scaling $\nu\propto N_{\BH}^{2} n_{\rm system}
v_{\vir}^{-1}$. Consider first galactic nuclei, hosting 2500 BHs and
approximately all of these BHs have undergone mergers \citep{pzm00}.
Galactic nuclei abundance in the universe is 100 times less than for
GCs (\S~\ref{sec:sub:GCabundance}). Assuming that the virial
velocity is a factor of $\sqrt{10}$ higher in galactic nuclei and
that the CO mass function has the same distribution as in GCs, we
get detection rates $\nu^{\gal}=5^{2}\times 1/100 \times 1/\sqrt{10}
\times \nu^{\gc}$. However, the large number of BH mergers likely
increases BH masses in galactic nuclei. For a uniform distribution
(i.e. $p=0$) of $N_{\BH}=2500$ between $m_{\min}=80$ and
$m_{\max}=100\Msun$ we get
$2500^{2}/[500\times(100-80)/(100-5)]^{2}\times
1/(100\sqrt{10})=1.8$ times the rates shown in the corresponding row
of Tab.~\ref{tab:models} for GCs: $4.\times 10^{-4}\yr^{-1}$ for
InLIGO, $2.1\times 10^{-3}$ for VIRGO, $0.61\yr^{-1}$ for AdLIGO,
and $5.2\times 10^{-6}\yr^{-1}$ for {\it LISA}. These numbers should
only be regarded as rough estimates, since they result from the
direct application of simplified GC model assumptions to galactic
nuclei. The calculation assumed uncorrelated two-body interactions
which does not hold for motion in the potential of galactic centers
\citep{rt96}.

\subsubsection{Primordial Black Holes in Galaxies}

For a second example consider the GW detections from galactic haloes
comprised of low-mass primordial BHs (PBHs) (see e.g.
\citealt{LIGO05b} and references therein). For a quick upper-limit
estimate on the PE detection rate we repeat our analysis for GCs by
changing the model parameters to describe galactic haloes. We assume
$N_{\rm PBH}=10^{11}$ PBHs within a maximum radius $R=5\,\rm kpc$, a
virial velocity $v_{\vir}=220\,{\rm km/s}$, and a uniform
distribution of masses between $m_{\min}=0.25\Msun$ and
$m_{\max}=0.95\Msun$. The maximum distance of a matched filtering
detection at a characteristic frequency of $0.9 f_{M \max}=4260\Hz$
with angular-averaged signal-to-noise ratio $S/N=5$ is 0.33, 0.62,
4.5, 0, and $0.06\Mpc$ for InLIGO, VIRGO, AdLIGO, {\it LISA}, and
{\it NGLISA}, respectively. The final result for the detection rate
after the compilation of the full analysis described in the paper
gives $\nu=1.5\times 10^{-11}$, $1.6\times 10^{-11}$, $2.9\times
10^{-10}$, $2.6\times 10^{-11}$, and $1.4\times 10^{-9}$,
respectively. These numbers are comparable to the total NS--NS PE
rate in GCs. It is 9 orders of magnitudes smaller than the event
rates estimates for PBH binary coalescence in one Milky Way sized
galaxy \citep{ioka98,LIGO05b}.

\subsubsection{Unresolved Parabolic Encounter Background}

Another extension of the present analysis is to estimate the number
of low $S/N$ PE events, adding an unresolved astrophysical
background to the GW detector noise budget similar to the unresolved
WD background \citep[e.g.][]{hbw90,nyp01,bdl04,cc05} and unresolved
capture sources \citep{bc04b}. Since PE rates are progressively
larger for progressively smaller characteristic frequencies, $f_0$,
and since all PE waveforms extend to GW frequencies $f\lsim f_0$, PE
background will be most substantial for space detectors, especially
around the minimum frequency noise wall
($f_{\min}=10^{-5}$--$10^{-4}\Hz$). The total number of PE events
within a distance $D$, can be obtained from (\ref{eq:dnu2})
neglecting cosmology as
\begin{equation}
R=\frac{4\pi}{3}D^3n^{\rm gc}\int_{f_{\min}}^{f_{\max}} \D\nu^{\rm
II} \approx 2\pi D^3n^{\rm gc}\nu^{\rm
II}_1\left(\frac{f_{\min}}{f_{100}}\right)^{-2/3}
\end{equation}
where $f_{\max}$ is the maximum frequency for PEs avoiding a
collision (\ref{eq:fMmax}), which drops out to leading order if
$f_{\min}\ll f_{\max}$. For $D=10\Gpc$ and $f_{\min}=10^{-5}\Hz$
$(10^{-4}\Hz)$ we get one event every $1/R=19\,{\rm sec}$ $(88\,{\rm
sec})$, which corresponds to $k=5300$ $(110)$ events for a
$10^{-5}\Hz$ ($10^{-4}\Hz$) frequency bin. If core contraction
enhances PE rates by a factor of $14^{4}$ (see
\S~\ref{sec:sub:CoreCollapse}), we get $k= 2\times 10^8$ $(4\times
10^6)$ events per frequency bin.

These events will typically have a very low signal-to-noise ratio,
e.g. for a single PE encounter with $f_0=10^{-5}\Hz$ for
$m_1=m_2=50\Msun$ at $\dL=10\Gpc$ we get $(S/N)_1\sim 8
\times10^{-9}$, and $f_0=10^{-4}\Hz$ yields $(S/N)_1\sim 2\times
10^{-7}$ for {\it LISA}. Assuming that average unresolved PE noise
accumulates proportional to $\sqrt{k}$, we get a net amplitude of
only $(S/N)_{\rm net}=10^{-4}$ or $5\times 10^{-4}$ for {\it LISA}
even in the core contracted case for frequencies $10^{-5}\Hz$ or
$10^{-4}\Hz$, respectively. Thus, we anticipate that the unresolved
PE background adds a negligible amount to the {\it LISA} noise, and
the unresolved PE background from stellar BH encounters will not be
an issue for near-future extensions either.

\acknowledgments We thank Kip Thorne for initial motivation and
suggestions. We thank Luca Matone, Zolt\'an Haiman, Kristen Menou,
Peter Shawhan, Patrick Sutton for valuable comments on the
manuscript, and the LIGO Scientific Collaboration Review, in
particular Vicky Kalogera for pointing out the relevance of Spitzer
instability for our estimates. We are grateful for the Caltech SURF
program and the LIGO collaboration for support. The authors
gratefully acknowledge the support of the United States National
Science Foundation and Columbia University in the City of New York.

\appendix
\section{Approximate Analytical Formulae}

The integrals given in Eq.~(\ref{eq:dnu2}) can be carried out
analytically as a Taylor-expansion with respect to the small
quantities $\kappa_{\BH}=m_{\min}/m_{\max}$ and
$\kappa_{\NS}=m_{\NS}/m_{\max}$. The PE event rate is calculated in
three parts
\begin{equation}
\nu^{\II}_1= \nu^{II}_{1,\BH-\BH} + \nu^{II}_{1,\BH-\NS} +
\nu^{\II}_{1,\NS-\NS},
\end{equation}
where $\nu^{II}_{1,\BH-\BH}$, $\nu^{II}_{1,\BH-\NS}$, and
$\nu^{II}_{1,\NS-\NS}$ are the event rates of BH-BH, BH-NS, and
NS-NS encounters. Substituting the mass dependence in
Eq.~(\ref{eq:dnu2}) we get,
\begin{eqnarray}\label{eq:app:nu2}
\nu^{\II}_1(m_1,m_2)&=& \frac{\G^{4/3}}{(4\pi)^{2/3}}
 \frac{N_{\rm CO}^2}{R_{\gc}^3 v_{\vir}}
 \frac{(m_1+m_2)^{4/3}m_>^{3/2}}{(m_1^{-1}+m_2^{-1})^{1/2}}
 f_{100}^{-2/3},\\
\nu^{\II}_2(m_1,m_2)&=& \frac{2^{1/3}\G^{2/3}}{3\pi^{4/3}}
 \frac{N_{\rm CO}^2 v_{\vir}}{R_{\gc}^3}
 (m_1+m_2)^{4/3}(m_1^{-1}+m_2^{-1})^{1/2}m_>^{3/2} f_{100}^{-4/3}.
\end{eqnarray}
We approximate the $g_{\NS}(m)$ distribution with a Dirac-$\delta$
function. Expanding the integrals in Taylor-series in $\kappa_{\BH}$
and $\kappa_{\NS}$, we get
\begin{eqnarray}
\nu^{\II}_{1,\BH-\BH} &=& 2^{-4/3} \frac{9}{187}\left( 2^{5/6} +
\frac{3}{5}\right)
 \left(1 + 2\kappa_{\BH} + 1.364\kappa_{\BH}^2\right)
 \left(\frac{\langle m_\BH \rangle}{\Msun}\right)^2
 \left(\frac{m_{\max}}{m_{\CO}}\right)^{4/3}\nu^{\I}_1,\label{eq:BH-BH}\\
\nu^{\II}_{1,\NS-\BH} &=& 2^{-1/3} \frac{3}{16}
 \left(1 + \kappa_{\BH} - \kappa_{\BH}^{4/3} +
 \frac{10}{3}\kappa_{\NS} +
 \frac{10}{3}\kappa_{\BH}^{1/3}\kappa_{\NS}\right)
 \left(\frac{\langle m_\BH \rangle}{m_{\CO}}\right)^{4/3}
 \left(\frac{m_{\max}}{\langle m_\BH \rangle}\right)^{1/3}
 \left(\frac{m_{\NS}}{\Msun}\right)^{2}
 \nu^{\I}_1,\nonumber\\\label{eq:BH-NS}\\
\nu^{\II}_{1,\NS-\NS} &=& 2^{-5/2}
 \left(\frac{m_{\CO}}{\Msun}\right)^{-4/3}
 \left(\frac{m_{\NS}}{\Msun}\right)^{3}
 \nu^{\I}_1,\label{eq:NS-NS}
\end{eqnarray}
where  $\nu^{\I}_1$ is the Model-I event rate given by
Eq.~(\ref{eq:nu1I}).

Next we present ready-to-use formulas for calculating the detection
rates of PEs. Assuming a constant density of GCs, no cosmological
and no general relativistic corrections, the detection rate per
logarithmic frequency bin becomes
\begin{equation}\label{eq:app:dnu/dlnf}
\frac{\D \nu^{\rm tot}}{\D \ln(f_0)} = k
\frac{(2\G\Msun)^{19/3}}{\cl^{12}} n^{\gc} N_{\CO}^2R_{\gc}^{-3}
v^{-1}_{\vir} f_0^{4/3}\left(\frac{W(f_0)}{S/N}\right)^3
K(f_0,\lambda)
\end{equation}
where $S/N$ is the minimum signal-to-noise ratio (which is set equal
to 5 in our numerical results above), $k\equiv
85^{3/2}2^{1/3}\pi^{7/3}/6144=2.323$ is a constant coefficient,
$v_{\vir}$ is the virial velocity (\ref{eq:v_vir}), $n^{\gc}$ is the
average GC density in the universe, $R_{\gc}$ is the typical radius
of the GC, $N_{\CO}$ is the number of COs in the GC,
$K(f_0,\lambda)$ and $W(f_0)$ are dimensionless terms,
$K(f_0,\lambda)$ depending on the CO mass distribution, $g_{\CO}$,
and $W(f_0)$ on the normalized GW energy spectrum $F(f/f_0)$
Eq.~(\ref{eq:h(f)2}), and the detector spectral noise density,
$S_n(f)$:
\begin{eqnarray}\label{eq:K(f_0)}
K(f_0,\lambda)&=&\int\int_{f_0\leq f_{0,\max}[\Msun(x_1+x_2),\lambda]}\D
x_1\D x_2\; g_{\CO}(x_1)g_{\CO}(x_2)\frac{(x_1 x_2)^{7/2}
x_{>}^{3/2}}{(x_1+x_2)^{1/6}}, \\
W(f_0)&=&\sqrt{\frac{4}{5}\int_{f_{\min}}^{f_{\max}}
\frac{1}{f^2}\frac{F(f/f_{0})}{S_n(f)^2}\;\D f}.
\end{eqnarray}
In terms of $f_0$, $K(f_0,\lambda)$ is constant for $f_0\leq
f_{0,max}(2m_{\min},\lambda)$, decreases monotonically for larger
$f_0$ and attains 0 for $f_0\geq f_{0,\max}(2m_{\max},\lambda)$ (see
Eq.~[\ref{eq:fconstraint}] for the definition of
$f_{0,\max}(M,\lambda)$, and $m_{\min}$ and $m_{\max}$ are the
minimum and maximum masses of the COs, respectively). In
Eq.~(\ref{eq:K(f_0)}), the integration variables $x_1$ and $x_2$ are
the dimensionless masses of the COs, for which $m_{\min}/\Msun \leq
x_{1,2} \leq m_{\max}/\Msun$. For core collapsed clusters $\Msun$
has to be changed to $m_{\rm core}$, the typical mass of individual
components in the core, $R_{\gc}$ has to be changed to $R_{\rm
core}$, and $v_{\vir}$ to $v_{\rm core}$. These values should be set
consistently with the core velocity dispersion and core radius which
are input parameters for a globular cluster model. For given $m$
components, the velocity dispersion is then $v_m=(m/m_{\rm
core})^{-1/2} v_{\rm core}$ and maximum radius from the cluster
center is $R_m=(m/m_{\rm core})^{-1/2} R_{\rm core}$.

The total detection rate of parabolic encounters (again assuming a
constant density of GCs and no cosmological and general relativistic
corrections) is
\begin{equation}\label{eq:app:nutot}
\nu^{\rm tot} = k \frac{(2\G m_{\rm
core})^{19/3}}{\cl^{12}}(S/N)^{-3} n^{\gc} N_{\CO}^2R_{\rm
core}^{-3} v^{-1}_{\rm core}
\int_{f_{\min}/10}^{f_{0,\max}(2M_{\max},\lambda)}\D f_0\,
f_0^{1/3}W(f_0)^3 K(f_0,\lambda).
\end{equation}


\begin{thebibliography}{84}

\bibitem[Abbott et. al.(2004)]{a04} Abbott, B. et al. 2004, Nucl. Instrum. Methods A, 517

\bibitem[Abbott et. al.(2005a)]{a05a} Abbott, B. et al. 2005, \prd, 72, 042002

\bibitem[Abbott et. al.(2005b)]{a05b} Abbott, B. et al. 2005, \prl, 94, 181103

\bibitem[Acernese et al.(2005)]{virgo05} Acernese, F. et al. 2005, Class. Quant. Grav., 22, S869


\bibitem[Ando et al.(2005)]{tama05} Ando, M. et al., the TAMA Collaboration, 2005, Class. Quant. Grav., 22, S881

\bibitem[Anninos et al.(1993)]{ann93} Anninos, P., Hobill, D., Seidel, E., Smarr, L., \& Suen W. 1993, \prl, 71, 182851


\bibitem[Baker et al.(2006)]{bak06} Baker, J. G., Centrella, J., Choi, D., \& van Meter, J. 2006, gr-qc/0602026

\bibitem[Barack \& Cutler(2004a)]{bc04a} Barack, L. \& Cutler, C. 2004a, \prd, 69, 082005

\bibitem[Barack \& Cutler(2004b)]{bc04b} Barack, L. \& Cutler, C. 2004b, \prd, 70, 122002

\bibitem[Binney \& Tremaine(1987)]{bt87} Binney, J., \& Tremaine, S. 1987, Galactic Dynamics (Princeton: Princeton Univ. Press)

\bibitem[Belczynski et al.(2005)]{belcz05} Belczynski, K., Sadowski, A., Rasio, F. A., \& Bulik, T. 2005, \apj, submitted, astro-ph/0508005


\bibitem[Benacquista, DeGoes, \& Lunder(2004)]{bdl04} Benacquista, M. J., DeGoes, J., \& Lunder, D. 2004, Class. Quant. Grav., 21, 509


\bibitem[Blanchet \& Sch\"{a}fer(1989)]{bs89} Blanchet, L. \& Sch\"{a}fer 1989, \mnras, 239, 845

\bibitem[Blanchet et al.(1995)]{bdiww95} Blanchet, L., Damour, T., Iyer, B. R., Will, C. M., \& Wiseman, A. G. 1995, \prl, 74, 3515

\bibitem[Blanchet et al.(2005)]{bdei05} Blanchet, L., Damour, T., Esposito-Far\`{e}se, \& Iyer, B. R. 2005, \prd, 71, 124004

\bibitem[Blecha et al.(2005)]{blecha05} Blecha L., Ivanovna, N., Kalogera, V., Belczynski, K., Fregeau, J., \& Rasio, F. 2005, \apj, submitted, astro-ph/0508597

\bibitem[Cardoso \& Lemos(2003)]{cl03} Cardoso, V. \& Lemos, J. P. 2003, \prd, 67, 08005

\bibitem[Casares(2005)]{cas05} Casares, J. 2005, The Many Scales of the Universe - JENAM 2004 Astrophysics Reviews, Kluwer Academic Publishers, astro-ph/0503071

\bibitem[Chandar, Whitmore, \& Lee(2004)]{cwl04} Chandar, R., Whitmore, B., Lee, M. G. 2004, \apj, 611, 220

\bibitem[Cutler, Kennefick, \& Poisson(1994)]{ckp94} Cutler, C., Kennefick, D., \& Poisson, E. 1994, \prd 50, 3816


\bibitem[Cornish \& Crowder(2005)]{cc05} Cornish, N. J. \& Crowder, J. 2005, \prd, 72, 043005

\bibitem[Crowder \& Cornish(2005)]{cc05b} Crowder, J. \& Cornish, N. J. 2005, \prd, 72, 083005

\bibitem[Cutler \& Thorne(2002)]{ct02} Cutler, C., \& Thorne, K. S. 2002, Proceedings of 16th international conference on General relativity and Gravitation (Eds N.T. Bishop and S.D. Maharaj) (2002); gr-qc/0204090

\bibitem[Davis et al.(1971)]{d71} Davis, M., Ruffini, R., Press, W. H., \& Price, R. P. 1971, \prl, 27, 1466

\bibitem[D'Eath \& Payne(1992)]{dp92} D'Eath, P. D., \& Payne, P. N., 1992, \prd, 46, 02694

\bibitem[DeSalvo (2004)]{desalvo04} DeSalvo, R. 2004, Class. Quant. Grav., 21, S1145

\bibitem[Djorgovski \& Meylan(1994)]{dm94} Djorgovski, S., \& Meylan, G. 1994, AJ, 108, 1292

\bibitem[Dymnikova, Popov, \& Zentsova(1982)]{dpz82} Dymnikova, I. G., Popov, A. K., \& Zentsova, A. S. 1982, Ap\&SS, 85, 231

\bibitem[Eisenstein(1997)]{eis97} Eisenstein, D. J. 1997, astro-ph/9709054

\bibitem[Farouki \& Salpeter(1982)]{fs82} Farouki, R. T., \& Salpeter, E. E. 1982, \apj, 253, 512

\bibitem[Flanagan \& Hughes(1998)]{fh98} Flanagan, \'{E}. \'{E}. \& Hughes, S. A. 1998, \prd, 57, 8

\bibitem[Freitag(2003)]{frei03} Freitag, M. 2003, \apj, 583, L21

\bibitem[Freitag, G\"{u}rkan, \& Rasio (2006)]{fgr06} Freitag, M., G\"{u}rkan, M. A., \& Rasio, F. A. 2006, \mnras, in press, astro-ph/0503130

\bibitem[Fryer \& Kalogera(2001)]{fk01} Fryer, C. L. \& Kalogera, V. 2001, \apj, 554, 548

\bibitem[Gair et al.(2004)]{gair04} Gair, J. R., Barack, L., Creighton, T., Cutler, C., Larson, S. L., Phinney, E. S., \& Vallisneri, M. 2004, Class. Quant. Grav., 21, 1595

\bibitem[Gair, Kennefick, \& Larson(2005)]{gkl05a} Gair, J. R., Kennefick, D. J., \& Larson, S. L. 2005, \prd, 72, 084009

\bibitem[Gair, Kennefick, \& Larson(2006)]{gkl05b} Gair, J. R., Kennefick, D. J., \& Larson, S. L. 2006, \apj, in press, astro-ph/0508275

\bibitem[Gebhardt, Rich, \& Ho(2005)]{grh05} Gebhardt K., Rich, R. M., \& Ho, L. C. 2005, \apj, 634, 1093



\bibitem[Grote et al.(2005)]{geo05} Grote, H. et al. 2005, Class. Quant. Grav., 22, S193

\bibitem[Goudfrooij et al.(2003)]{gsbkme03} Goudfrooij, P., Strader, J., Brenneman, L., Kissler-Patig, M., Minniti, D., \& Edwin Huizinga, J. 2001, \mnras, 343, 665

\bibitem[G\"{u}ltekin, Miller, \& Hamilton(2006)]{gmh05} G\"{u}ltekin, K., Miller, M. C., \& Hamilton, D. P 2006, \apj, in press, astro-ph/0509885

\bibitem[G\"{u}rkan, Freitag, \& Rasio(2004)]{gfr04} G\"{u}rkan, M. A., Freitag, M., \& Rasio, F. A. 2004, \apj, 604, 632

\bibitem[Heggie, Trenti, \& Hut(2006)]{hth06} Heggie, D. C., Trenti, M., \& Hut, P. 2006, astro-ph/0602408

\bibitem[Hills \& Day(1976)]{hd76} Hills, J. G., \& Day, C. A. 1976, Astrophys. Lett., 17, 87


\bibitem[Hils, Bender, \& Webbink(1990)]{hbw90} Hils, D., Bender, P. L., \& Webbink, R. F. 1990, \apj, 360, 75

\bibitem[Hopman \& Alexander(2005)]{ha05} Hopman, C., \& Alexander, T. 2005, \apj, 629, 362

\bibitem[Hopman \& Portegies Zwart(2005)]{hp05} Hopman, C., \& Portegies Zwart, S. 2005, \mnras, 363, L56

\bibitem[Hughes et al.(2001)]{hmbh01} Hughes, S. A., M\'{a}rka, S., Bender, P. L., \& Hogan, C. J. 2001, Proc. of the APS/DPF/DPB Summer Study on the Future of Particle Physics (Snowmass 2001) ed. Graf, N. eConf, C010630, 402, astro-ph/0110349


\bibitem[Ioka et al.(1998)]{ioka98} Ioka, K., Chiba, T., Tanaka, T., \& Nakamura, T. 1998, \prd, 58, 063003

\bibitem[Jaronowski et al.(1996)]{jkkt96} Jaronowski, P., Kokkotas, K. D., Kr\'{o}lak, A., \& Tsegas, G. 1996, Class. Quantum Grav., 13, 1279

\bibitem[Kalogera et al.(2004)]{kal04} Kalogera, V. et al. 2004, \apjl, 601, L179

\bibitem[Khanna et al.(1999)]{khanna99} Khanna, G., Baker, J., Gleiser, R. J., Laguna, P., Nicasio, C. O., Nollert, H., Price, R., \& Pullin, J. 1999, \prl, 83, 3581

\bibitem[Kocsis \& G\'{a}sp\'{a}r(2004)]{kg04} Kocsis, B. \& G\'{a}sp\'{a}r, M. E. 2004, LIGO note, http://www.ligo.caltech.edu/docs/T/T030136-00.pdf

\bibitem[Kov\'{a}cs \& Thorne(1978)]{kt78} Kov\'{a}cs, S. J., \& Thorne, K. S. 1978, \apj, 224, 62

\bibitem[Kroupa \& Weidner(2003)]{kw03} Kroupa, P., \& Weidner, C. 2003, \apj, 598, 1076

\bibitem[Lazzarini \& Weiss(1996)]{lw96} Lazzarini, A., \& Weiss, R. 1996, LIGO note, http://www.ligo.caltech.edu/docs/E/E950018-02.pdf

\bibitem[Lee(1993)]{lee93} Lee, M. H. 1993, \apj, 418, 147

\bibitem[LIGO Scientific Collaboration (2005a)]{LIGO05a} LIGO Scientific Collaboration, 2005a, \prd, 72, 062001

\bibitem[LIGO Scientific Collaboration (2005b)]{LIGO05b} LIGO Scientific Collaboration, 2005b, \prd, 72, 082002

\bibitem[Lipunov, Postnov, \& Prokhorov(1997)]{lpp97} Lipunov, V. M., Postnov, K. A., \& Prokhorov, M. E. 1997, NewA, 2, 43

\bibitem[Lotz, Miller, \& Fergusson(2004)]{lmf04} Lotz, J. M., Miller, B. W., Ferguson, H. C. 2004, \apj, 613, 262

\bibitem[Martel(2004)]{mar04} Martel, K. 2004, \prd, 69, 044025

\bibitem[Meylan(1987)]{mey87} Meylan, G. 1987, A\&A, 184, 144

\bibitem[Mik\'{o}czi, Vas\'{u}th, \& Gergely(2005)]{mvg05} Mik\'{o}czi, B., Vas\'{u}th, M., \& Gergely, L. \'{A}. 2005, \prd, 71, 124043

\bibitem[Miller(2002)]{mil02} Miller, M. C. 2002, \apj, 581, 438

\bibitem[Miller \& Colbert(2004)]{mc04} Miller, M. C. \& Colbert, E. J. M. 2004, Int'l J. of Mod. Phys. D, 13, 1, 1

\bibitem[Misner, Thorne, \& Wheeler(1973)]{Gravitation} Misner, C. W., Thorne, K. S., \& Wheeler, J. A. 1973, Gravitation (San Francisco: Freeman)

\bibitem[Mine, Shibata, \& Tanaka(1996)]{mst96} Mine, Y., Shibata, M., \& Tanaka, T. 1996, \prd, 53, 020622

\bibitem[Moreschi(1999)]{mor99} Moreschi, O. M. 1999, \prd, 59, 084018

\bibitem[Nakar, Gal-Yam, \& Fox(2005)]{ngf05} Nakar, E., Gal-Yam, A., \& Fox, D. B. 2005, \apj, submitted, astro-ph/0511254

\bibitem[Nelemans, Yungelson, \& Portegies Zwart(2001)]{nyp01} Nelemans, G., Yungelson, L. R., \& Portegies Zwart, S. F. 2001, A\&A, 375, 890

\bibitem[O'Leary et al.(2006)]{O'Leary05} O'Leary, R. M., Rasio, F. A., Fregeau, J. M., Ivanova, N., \& O'Shaughnessy, R. 2006, \apj, 637, 937



\bibitem[Portegies Zwart et al.(2006)]{pz06} Portegies Zwart, S. F., Baumgardt, H., McMillan, S. L. W., Makino, J., Hut P., \& Ebisuzaki, T. 2006, \apj, in press, astro-ph/0511397

\bibitem[Portegies Zwart \& McMillan(2000)]{pzm00} Portegies Zwart, S. F. \& McMillan, S. L. W. 2000, \apj, 528, 17

\bibitem[Portegies Zwart \& Spreeuw(1996)]{ps96} Portegies Zwart, S. F. \& Spreeuw, H. N. 1996, A\&A, 312, 670

\bibitem[Portegies Zwart \& Yungelson(1998)]{py98} Portegies Zwart, S. F. \& Yungelson, L. R. 1998, A{\&}A, 312, 670


\bibitem[Postnov \& Prokhorov(2003)]{pp01} Postnov, K. A. \& Prokhorov, M. E. 2001, Astr. Rep., 45, 899

\bibitem[Price \& Pullin(1994)]{pp94} Price, R. H., \& Pullin, J. 1994, \prl, 72, 3297

\bibitem[Pryor \& Meylan(1993)]{pm93} Pryor, C., \& Meylan, G. 1993, in ASP Conf. Ser. 50, Structire and Dynamics of Globular Clusters, ed. S. G. Djorgovski \& G. Meylan (San Fransisco: ASP), 357

\bibitem[Ptak \& Colbert(2004)]{pc04} Ptak, A., \& Colbert, E. 2004, \apj, 606, 291

\bibitem[Rauch \& Tremaine(1996)]{rt96} Rauch, K. P., \& Tremaine, S. 1996, New Astronomy, 149

\bibitem[Saslaw(1985)]{Saslaw} Saslaw, Gravitational physics of stellar and galactical systems, 1985, Cambridge Univ. Press, 52

\bibitem[Sasaki \& Nakamura(1982)]{sn82} Sasaki, M. \& Nakamura, T. 1982, Prog. Tbeor. Phys., 67, 1788

\bibitem[Scalo(1986)]{Scalo86} Scalo, J. M. 1986, Fundam. Cosmic Phys., 11, 1

\bibitem[Seto, Kawamura, \& Nakamura (2001)]{DECIGO} Seto, N., Kawamura, S., Nakamura, T. 2001, \prl, 87, 221103


\bibitem[Sigurdsson \& Hernquist(1993)]{sh93} Sigurdsson, S., \& Hernquist, J. 1993, \nat, 364, 423

\bibitem[Sigurdsson \& Phinney(1995)]{sp95} Sigurdsson, S., \& Phinney, E. S. 1995, \apjs, 99, 609

\bibitem[Sigurdsson \& Rees(1997)]{sr97} Sigurdsson S., \& Rees, M. J. 1997, \mnras, 284, 318

\bibitem[Sperhake et al.(2005)]{sper05} Sperhake, U., Kelly, B., Laguna, P., Smith, K. L., \& Schnetter E. 2005, \prd, 71, 124042

\bibitem[Spitzer (1969)]{spitz69} Spitzer, L., Jr. 1969, \apj, 158, L139

\bibitem[van den Bergh(2005)]{bergh05} van den Bergh, S. 2005, \apj, submitted, astro-ph/0509811



\bibitem[Tegmark et al.(2004)]{teg04} Tegmark, M. et al. 2004, \prd, 69, 103501


\bibitem[Timpano, Rubbo, \& Cornish(2005)]{trc05} Timpano, S., Rubbo, L. J., \& Cornish,  N. J. 2005, gr-qc/0504071

\bibitem[Thorne(1987)]{thorne87} Thorne, K. S. 1987, in 300 Years of Gravitation, ed. Hawking, S. W. \& Israel, W., Cambridge University Press, pp. 330 -- 458

\bibitem[Tully(1988)]{Tul88} Tully, R. B. 1988, \aj, 96, 73

\bibitem[Turner(1977)]{Turner} Turner, M. 1977, \apj, 216, 610


\bibitem[Watters, Joshi, \& Rasio(2000)]{wjr00} Watters, W. A., Joshi, K. J., \& Rasio, F. A. 2000, \apj, 539, 331

\bibitem[Will(2004)]{w04} Will, C. M. 2004, \apj, 611, 1080



\end{thebibliography}
\end{document}